\documentclass[aps,pre,groupedaddress,amssymb,amsmath,pra,floatfix,showpacs]{revtex4}
\usepackage{epsfig}

\begin{document}
\title{Small Phase Space Structures and their Relevance to Pulsed Quantum Evolution:\\
the Stepwise Ionization of the Excited Hydrogen Atom in a Microwave  Pulse}
\author{Luca C. Perotti}
\affiliation{Department of Physics, Texas Southern University, Houston, Texas 77004 USA}
\date{\today}
\begin{abstract}

Experiments have shown that the microwave ionization probability of a highly excited almost monodimensional
hydrogen atom subjected to a microwave pulse sometimes grows in steps when the peak electric field of the pulse is
increased. Classical pulsed simulations display the same steps, which have been traced to phase-space metamorphoses.
Quantum numerical calculations again exhibit the same ionization steps. I show that the time-sequence of two level
interactions, responsible for the observed steps in the quantum picture, is strictly related to the classical phase
space structures generated by the above mentioned metamorphoses.
\end{abstract}
\pacs{05.45.Mt, 03.65.-w, 32.80.Rm, 47.20.Ky}
\maketitle


\section{Introduction}

In classical physics, integrability - even for nondissipative systems - is the exception \cite{ref1}. Regular -
both periodic and quasiperiodic - and chaotic classes of motion can coexist in phase space. Classes of motion can
appear or disappear when the system parameters are changed; periodic orbit bifurcation points and thresholds for
chaos are among the most widely known of the critical points where this happens.

The relevance of nonintegrable classical dynamics to quantum
dynamics has been \cite{ref2} and still is the source of much
discussion, which has lately concentrated mainly on three problems:
dynamical tunneling, both chaotic or resonance assisted \cite{tun},
the construction of atomic nondispersive wave packets \cite{nond},
and the ionization of Rydberg atoms \cite{atom}.

Existing semiclassical theories predict a direct dependance of the quantum wavefunction on the corresponding classical physics,
for instance via the classical action function (when it can be built). But the temporal regions of applicability of
such theories generally are not well delineated, especially near the critical values and for chaos. Little is known
about the case where a parameter is {\it swept in time} through a classical critical value. Under what conditions
and to what extent can the evolution of a nonintegrable quantum system be ``guided" by short-time nonadiabatic
behavior present in its corresponding nonintegrable classical evolution?

One semiclassical dynamical system which has been for a long time the object of
extensive experimental study is a highly excited hydrogen atom
exposed to a partially ionizing short pulse of microwave electric
field. If the atom is prepared in a static electric field collinear
with the microwave field, and an extreme Stark state is excited, a
one-dimensional approximation can be used, thus greately
simplifying the treatment. Experimental and quantum numerical
results are often found to be nearly classical, agreeing reasonably
well with numerical predictions based on classical models. The
classical electron dynamics is a microwave frequency sensitive
mixture of regular motion, locally chaotic motion, and globally
chaotic motion \cite{ref3}. The last motion eventually results in
ionization of the classical atom. Resonant regular motion is
displayed as island features in stroboscopic surfaces of section of
the phase space.

Let the initial principal quantum number of the atom be $n_0$. When the ratio $\omega_0
\equiv n_0^3 \omega$ of the microwave frequency $\omega$ to the
initial Kepler orbit frequency $1/n_0^3$ is near unity, resonant
classical motion dominates the character of a number of
semiclassical Floquet eigenstates which are experimentally populated
at the peak of the pulse \cite{ref3}. This inhibition of chaos
produces an experimentally observed increase in the threshold field
for ionization in the quantum system \cite{ref4}. Bounding the
island region in phase space is a locally chaotic region which can
be crossed by the system during the rise and fall of the pulse
\cite{ref5}. Such crossings are examples of a phase-space transport
process which would not occur in a fixed-field experiment
\cite{ref6,ref42}. They provide an explanation for observed nearly
classical final quantum state distributions which are double peaked
as a function of the principal quantum number \cite{ref7}. I shall
call {\it principal primary resonance zone} the $\omega_0 = 1$
resonance island together with the surrounding band of local chaos.

For different constant values of the microwave field strength, the
structure of  the resonance zones can be different. The structural
changes have been called phase-space metamorphoses \cite{ref8}, and
it has been suggested that they could influence the microwave
ionization process. For example, in Ref. \cite{ref8} it was shown
that the destruction of the last invariant curve between a primary
resonant zone and the global chaos above it can lengthen the classical
electron's escape time for ensembles of initial conditions in the
global chaos region. Here I am interested in the metamorphoses of
the internal structure of the resonance islands themselves. These
metamorphoses are associated with periodic orbit bifurcations which
generate, within the primary resonance zones, new ``secondary"
resonance Kolmogorov-Arnold-Moser (KAM) island chains, and with the
destruction of these same secondary resonance island chains.

Laboratory observations have been reported of sequences of steps in
the (experimentally defined) ionization of almost monodimensional
Hydrogen Rydberg atoms versus the peak field strength of the
microwave pulse \cite{ref9}. Classical numerical simulations
reproduce such steps. These classical step have been shown to be
due to metamorphosis-induced variations of the probability for the
transport in phase space responsible for ionization \cite{ref9}.
Together with more extensive numerical evidence and a more detailed
analysis of the classical process \cite{ref40}, I now present an
explanation of the good agreement of the quantum simulations with
the classical ones: even if, in terms of Floquet states, the
quantum evolution is dominated by a sequence of two level processes
(avoided crossings traversed by the system) and can therefore be
viewed as deeply quantum, the interactions between these couples of
states exist because of the classical secondary island chains.

The present paper is organized as follows: in section II I present the
system and the numerical techniques I used to investigate it; the results
of my investigation are presented in section III. Section IV is dedicated
to the interpretation of these results; section V finally sums up my
findings.

\section{The system and numerical methods}

The system I investigate is a $1D$  model for a stretched highly excited
hydrogen atom in collinear static and monochromatic microwave electric
fields \cite{ref10}. In atomic units the Hamiltonian reads
\begin{equation}
H = {\frac {p^2} 2} - {\frac 1 z} +z[F(t) sin{( \omega t +
\phi_0)}
- F_s], \hspace{.3in} z \geq 0,
\label{eq1}
\end{equation}
where $F(t)$ is the strength of the microwave field, $\phi_0$ is
its initial phase, and $F_s$ is the static field strength. To
simulate the experimental situation described in Ref. \cite{ref9},
the envelope of the microwave pulse is chosen to be
\begin{eqnarray}
F(t) = F^{max} \sin {(\pi t/T_p)},\nonumber
\end{eqnarray}
where $T_p = 7.78 \hspace{.1in}ns$ is the length of the microwave
pulse. For most of the simulations in the present paper this
corresponds to $136$ microwave periods.

As is costumary for this system I shall use classically
scaled parameters \cite{ref10}: $F_0^{max}= n_0^4 F^{max}$,
$F_{s0}= n_0^4 F_s$, and $\omega_0 = n_0^3 \omega$.
The scaled frequency is furthermore corrected to the first order in $F_{s0}$
to compensate for the Stark shift of the atomic frequencies \cite{ref3}:
\begin{eqnarray}
\omega_0 ' =  \omega_0 /(1 - 3F_{s0}).\nonumber
\end{eqnarray}

\subsection{Classical methods}

Following Ref. \cite{ref14}, classical numerical integration in
Ref. \cite{ref40} was performed in free-atom action angle variables
$(I,\theta)$, valid when the electron's energy is negative and
defined by the equations:
\begin{eqnarray}
z = 2I^2 \sin^2 (\xi /2),\nonumber\\ p = (1/I) \cot{(\xi/2)},\nonumber
\end{eqnarray}
where the eccentric anomaly angle $\xi (\theta)$ is defined by $\theta \equiv \xi - \sin{\xi}$.

To avoid equations of motion  containing terms which diverge as $z$
approaches zero, a dummy time $\eta$, defined by $dt \equiv (1 -
\cos{\xi})d \eta$ was also introduced \cite{ref12}. The true time t
increases monotonically with $\eta$. Newton's equations of motion
then become
\begin{eqnarray}
dI/d\eta = -I^2 \sin{\xi} [F(t) \sin{(\omega t + \phi_0)} -
F_s],\hspace{.7in}\nonumber\\ d\xi/d\eta = 1/I^3
+2I(1-\cos{\xi})[F(t)
\sin{(\omega t + \phi_0)} - F_s],\nonumber\\
dt/d\eta = 1 -\cos{\xi}.\hspace{2.2in}\nonumber
\end{eqnarray}
The above three ordinary differential equations were numerically integrated
using a fixed step, fourth order Runge-Kutta routine. The
ensemble of initial conditions $(\theta_0,I_0)$ was chosen so that, to first
order in $F_s$, the (classical) electron energy $E (t=0)$ be equal to the value of
the energy of the experimentally prepared initial quantum energy eigenstate:
\begin{equation}
-{\frac 1 {2n_0^2}} - {\frac 3 2}F_s n_0^2 = E(0) = -{\frac 1 {2I_0^2}} - F_sI_0^2(1 - \cos{\xi_0}).
\label{eq3}
\end{equation}
Given a value of $\theta_0(\xi_0)$, this equation determines a
corresponding value of $I_0$.

Classical values for the ``ionization" probability $P_I$ at the end of the pulse were averaged over uniform
distributions of the initial angle $\theta_0$ and microwave field phase $\phi_0$. The ``ionization" probability
contains two contributions. One comes from trajectories which are terminated at some time during the pulse where the
instantaneous value of the energy $E$ exceeds the value $-2\sqrt{F_s}$ for rapid ionization in the static field
alone. To match the experimental definition of ``ionization" \cite{ref9}, the other contribution comes from
trajectories whose final value of $E$ corresponds, according to equation (\ref{eq3}) above, to energy eigenstates
with quantum numbers outside the interval $n \in [50,90]$.

Stroboscopic surfaces of section of the long-time evolution in
$(\theta,I)$ space were also computed for $F(t)$ equal to various
constants $F$ \cite{ref40}. These reveal the long-time structures
existing in phase space for the various values of $F$
instantaneously traversed during the microwave pulse. For a
constant amplitude microwave field, the action-angle generalized
phase space (in the three coordinates $I$, $\theta$, and $t$) is a
well known \cite{ref3} admixture of zones of regular and chaotic
motion. Zones of regular electron motion include those occupied by
nested vortex tubes and those of the field-modulated atom
\cite{ref5}. Chaotic zones include zones of local chaos and of
global chaos. Stroboscopic surfaces of section (Poincar\'{e} maps)
reflect these zones in the full phase space as characteristic zones
in the action-angle $(I, \theta)$ subspaces (``sectioned" phase
space): the vortex tubes produce resonance island zones and the
local chaos produces zones of irregular motion which have been
called ``separatrix" zones \cite{ref13} (the actual separatrix of
integrable systems is ``broken" in the nonintegrable driven
hydrogen atom problem and becomes a chaotic layer which grows with
increasing microwave field strength). These structures are
self-similar: secondary resonance zones are found within primary
resonance zones and so on; but while all primary resonance zones
are born at zero microwave field, periodic orbit bifurcations
produce secondary resonance zones at nonzero field values.

The comparison of instantaneous ensemble distributions in phase
space with these long time surfaces of section, for various times
$t$ during the pulse \cite{ref9} has proven a useful tool for the
understanding of the evolution of the ensemble itself: when a phase
space structure influences the pulsed evolution, the ensemble
appears deformed by it in a characteristic way. For example, an
ensemble of points initially on a straight line is stretched by a
resonance island into a whorl \cite{ref43} as the points of the
ensemble move along the island's invariant curves. Local whorl-like
deformations of the ensamble are therefore indication of the
influence of an underlying island which can be identified with the
help of the surface of section at that time. Similarly, local
hyperbolic chaos deforms the ensemble into tendrils \cite{ref43}.

\subsection{Quantum methods}

The time evolution of the quantum system is evaluated by numerical integration of
the Schrodinger equation with $H$ given by eq.(\ref{eq1}) on a finite
subset of the (bound) free atom basis $\psi_n(z)$ chosen big enough so that
the probability reflected by the borders is small \cite{ref14}. Given an
initial state $\psi(z,0) = \Sigma_n C_n(0) \psi_n(z)$, the equation for the
evolution of the expansion coefficients $C_n(t)$ is
\begin{eqnarray}
i {\frac {dC_n(t)} {dt}} = E_n C_n(t) + {\mathcal F}(t)
\Sigma_m Z_{n,m} C_m(t), \nonumber\\ {\mathcal F}(t) = (F(t)
sin(\omega t +
\phi_0) - Fs),
\label{eq4}
\end{eqnarray}
where $Z_{n,m}$ is the matrix element of the operator $z$ between
the states $n$ and $m$. We approximate ${\mathcal F} (t)$ with the
function ${\mathcal F'}(t) = {\mathcal F}(t)\Delta t \Sigma_k
\delta(t - k\Delta t)$ which tends to ${\mathcal F} (t)$ for  $\Delta
t \to 0$ (in the sense that the integral of their difference over
an arbitrary time interval goes to zero as $\Delta t$)
\cite{ref14}. This procedure involves unphysical parameters which
have to be carefully chosen as not to falsify the results of the
integration: I discuss my choices in Appendix \ref{app:uno}.

The initial state for our simulations is taken to be the eigenstate with quantum number $n_0$
of the atomic Hamiltonian in the static field alone, calculated in the chosen
finite subset of the free atom basis.

\subsubsection{Quasienergy curves}

Like in the classical case, an understanding of the dynamics at
constant microwave amplitude will help illuminating the pulsed
dynamics. If $F(t)$ is a constant, Schrodinger's equation is a
differential equation with time-periodic coefficients. Floquet's
theory is therefore applicable \cite{ref15} and tells us that the
equation has solutions in the form
\begin{eqnarray}
\psi_i (t) = \Phi_i (t) e^{-i\varepsilon_i t/ \hbar}, \hspace{.4in} \Phi_i (t + T)= \Phi_i (t)
\nonumber
\end{eqnarray}
where $T$ is the period of the microwave, the constants
$\varepsilon_i$ take the name of quasienergies and the functions
$\psi_i (t)$  are called quasienergy (or Floquet) states. For
periodic systems quasienergies take the place of energies in the
description of the dynamical properties of the system. It can be
easily seen  that the same physical state $\psi_i (t)$ can be
written as
\begin{eqnarray}
\psi_i (t) = [\Phi_i (t) e^{-i2 \pi nt/T} ]e^{-i\varepsilon_i t/\hbar+i2\pi nt/T}\nonumber
\end{eqnarray}
where $\Phi_i (t) e^{-i2\pi nt/T}$ is again a T-periodic function. We
therefore have an infinity of replicas of the quasienergy spectrum
shifted by  $2\pi \hbar /T$ one respect to another. It will then be
sufficient to restrict ourselves to the energy interval $[0, 2\pi \hbar /T)$
(first Brillouin zone) to have all the levels. In particular, since the
free atom has a continuum, this too will be brought into the first Brillouin
zone. Due to the presence of the free atom continuum, the microwave
transforms all the free atom states into resonances with finite widths $\Gamma_i$: the
quasienergy spectrum for an atom in microwaves is therefore absolutely
continuous (no point spectrum exists) \cite{ref16}.

For the actual calculation of the quasienergies it is useful to
resort to the time evolution operator $G$ over one period $T$ of
the microwave (Floquet operator). There isn't just one Floquet
operator, but a whole family of them, parametrized by the phase
$\phi= (\omega t_0 + \phi_0)$ of the field at the beginning of the
period. While the eigenfunctions $\Phi_i (\phi)$ of each member of
the family are different and represent the different spatial
structures of the states at different times during the microwave
period, the family shares the same eigenvalues $G_i$. The
eigenvalues $G_i$ of the Floquet operator and the quasienergies are
connected by the relationship $G_i = e^{-i\varepsilon_i T/\hbar}$.
One common procedure to obtain the quasienergies is therefore to
numerically calculate a one period evolution operator and
diagonalize it thus obtaining the $G_i$s. The above relationship
then gives us the quasienergies on the circle $[0, 2\pi \hbar /T)$.

To investigate the evolution of a pulsed system, and also -as we
shall see in Appendix \ref{app:uno}- to control numerical errors
and approximations, it is useful to plot the quasienergies as
functions of the microwave field strength $F$. The resulting {\it
quasienergy curves} (see e.g. Fig. \ref{fig1}) help us connect the
characteristics of the system at different values of the microwave
field strength. The quasienergies change with $F$ and undergo a
sequence of avoided crossings of different widths. Since in the
$1D$ hydrogen atom all the quasienergies fall into the same
symmetry class \cite{ref17}, the Von Neumann-Wigner theorem
\cite{ref18} tells us that they all repel each other; this implies
the absence of real level crossings. Just like in the simulations
of the pulsed system, the Floquet operator is calculated on a
truncated basis and some caution has to be exerted to avoid
spurious results. A study of the quasienergy curves themselves can
help us in deciding what parts of what curves to trust
\cite{ref19}: see Appendix \ref{app:uno}.

\subsubsection{Quantum nonlinear resonances}

\begin{figure}[htbp]
\centering\epsfig{file=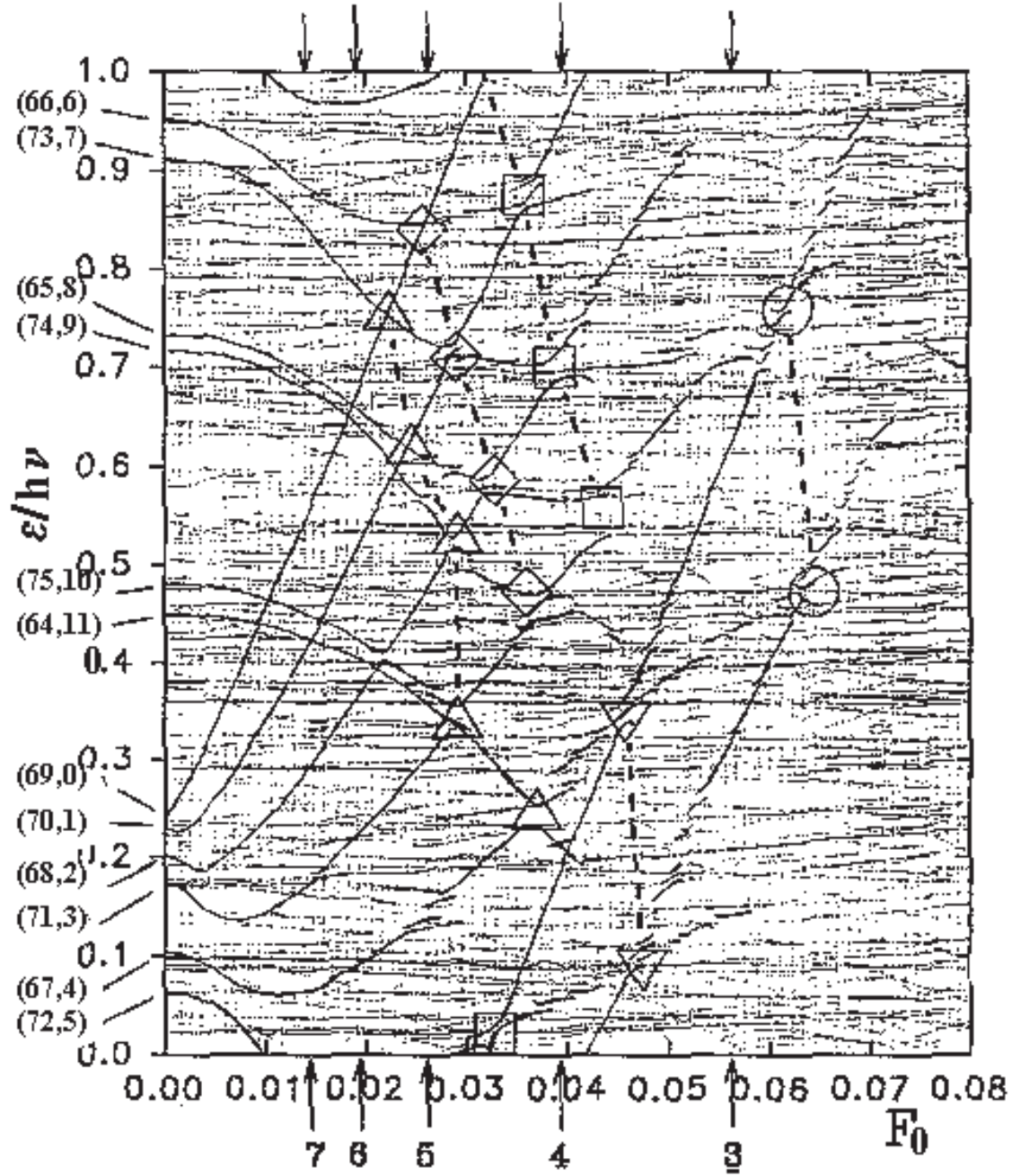,width=1.0\linewidth}
\caption{Quasienergy curves for $n = 17.50 GHz$ and $F_s = 8 V/cm$
calculated on the basis $n\in[30,221]$. The horizontal axis is the
microwave field strength rescaled for $n=69$. The darker curves are
those related to the $s/r =1$ quantum resonance. The labels $(n,k)$
give the free atom quantum number $n$ and the resonance quantum
number $k$. The groups of avoided crossings related to the most
visible classical secondary island chains with rotation number
$1/q$ are indicated by the following symbols: $q=3$ circles; $q=4$
down triangles; $q=5$ squares; $q=6$ diamonds; $q=7$ up triangles.
To aid the eye, lines are passed through the members of each group.
Also, arrows mark the classical numerical bifurcation fields,
labeled by $q$. (From Ref. \cite{ref40})}
\label{fig1}
\end{figure}

The behaviour of a grouping of levels strongly interacting
(repelling each other) at or close to zero microwave field is
particularly evident in Fig. \ref{fig1}. I have marked them as darker
lines. If we look at the (zero microwave field) {\it energies} of
these levels we see that each level is {\it close to a quantum
resonance}
\begin{equation}
\Delta E \equiv  E_{n+r} - E_n = s \omega \hbar
\label{eq7}
\end{equation}
with any of the other levels of the same grouping; the ratio $s/r$
being equal to $1$ for any two levels of the grouping. This
grouping represents on the $(E,F)$ plane what Ref. \cite{ref20}
calls a nonlinear quantum resonance: a finite (because of level
anharmonicity) number of quantum states whose interaction is due to
the existence of a (primary) classical resonance zone, in this case
the $\omega'_0 =1$ resonance zone. The structure these levels form
can be described as an excitation of the system induced by the
perturbation: Ref. \cite{ref45} calls it a ``Floton". The state of
the Floton is decribed by the new quantum number $k$ given in Fig.
\ref{fig1} next to the free atom one $n$; a simple interpretation of it in
terms of pendulum approximation is given in Appendix \ref{app:dueb}. The most
stable state is the ground one $(k=0)$; the excited states
$k=1,2,..$ are progressively shorter lived \cite{ref23}, as long as
we consider only the properly resonant states: those for which the
inflection point of the corresponding quasienergy curve in Fig. \ref{fig1}
lies to the left of the considered value of the microwave field
strength $F$ (see Appendix \ref{app:dueb}).

\subsubsection{Husimi function}

One of the most widely used representations of a wavefunction in
phase space is the Husimi function \cite{ref21}. Given a quantum
function $\psi(t)$, its Husimi function at a point
$\{\left<{p}\right>,\left<{q}\right>\}$ in phase space is the
projection (in atomic units)
$\rho_H(\left<{p}\right>,\left<{q}\right>,t) = | \left<{
\phi_{\left<{p}\right>\left<{q}\right>} | \psi(t) }\right> |^2$ of
$\psi(t)$ on a (minimum uncertainty) packet
$\phi_{\left<{p}\right>\left<{q}\right>}$ centered on that same
point in phase space and called the ``coarse graining function". In
action-angle space, due to the periodicity in the angle variable,
the standard (position-momentum space) choice of coarse graining
function \cite{ref21} is not possible. Following Ref. \cite{ref22}
I therefore take
\begin{eqnarray}
\phi_{\left<{I}\right>\left<{\theta}\right>}=\Sigma_{n=0}^{\infty}[{{\frac {\alpha^{(\alpha n+1)}}
{2\pi \Gamma (\alpha n+1)}} \left<{I}\right>^{\alpha n} e^{-\alpha
\left<{I}\right>}}]^{1/2} e^{i2\pi \left<{\theta}\right> n}\psi_n.\nonumber
\end{eqnarray}
Tests with different values of the width parameter $\alpha$ have
given - both for Ref. \cite{ref22} and for me - Husimi functions
having essentially the same shape when $\alpha < n_0$. For $n_0
\gg 1$ the classical zones in phase space are reflected in the character
of the Floquet eigenstates (FE), especially as seen through their
Husimi functions (see refs. \cite{ref19,ref22} and - for
semiclassical FE - Ref. \cite{ref5}). Thus regular FE can be either
resonance island FE or modulated atom FE. Irregular FE include the
separatrix FE, possibly ``scarred" by a high electron probability
density in phase space along an unstable periodic orbit
\cite{ref8,ref23}; in sectioned phase space this corresponds to a
maximum of the Husimi function on the unstable fixed point which
(surprisingly) has a simple interpretation in pendulum
approximation \cite{ref45}. At intermediate values of microwave
field strength, these scarred separatrix FE exhibit partial quantum
localization along the unstable periodic orbits (rather than full
localization on nested vortex tubes), and otherwise are distributed
within a bounded region of phase space classically occupied by
chaotic trajectories \cite{ref5,ref23}. The comparison of Husimi
functions for the quasienergy states with the classical surfaces of
section at the same values of $\omega$, $F$ and $\phi_0$ has often
proven quite instructive. Comparison of Husimi functions and
classical instantaneous ensembles for the pulsed simulations is
also possible, but I found them less useful in the present study.

\section{Results}

\begin{figure}[htbp]
\centering\epsfig{file=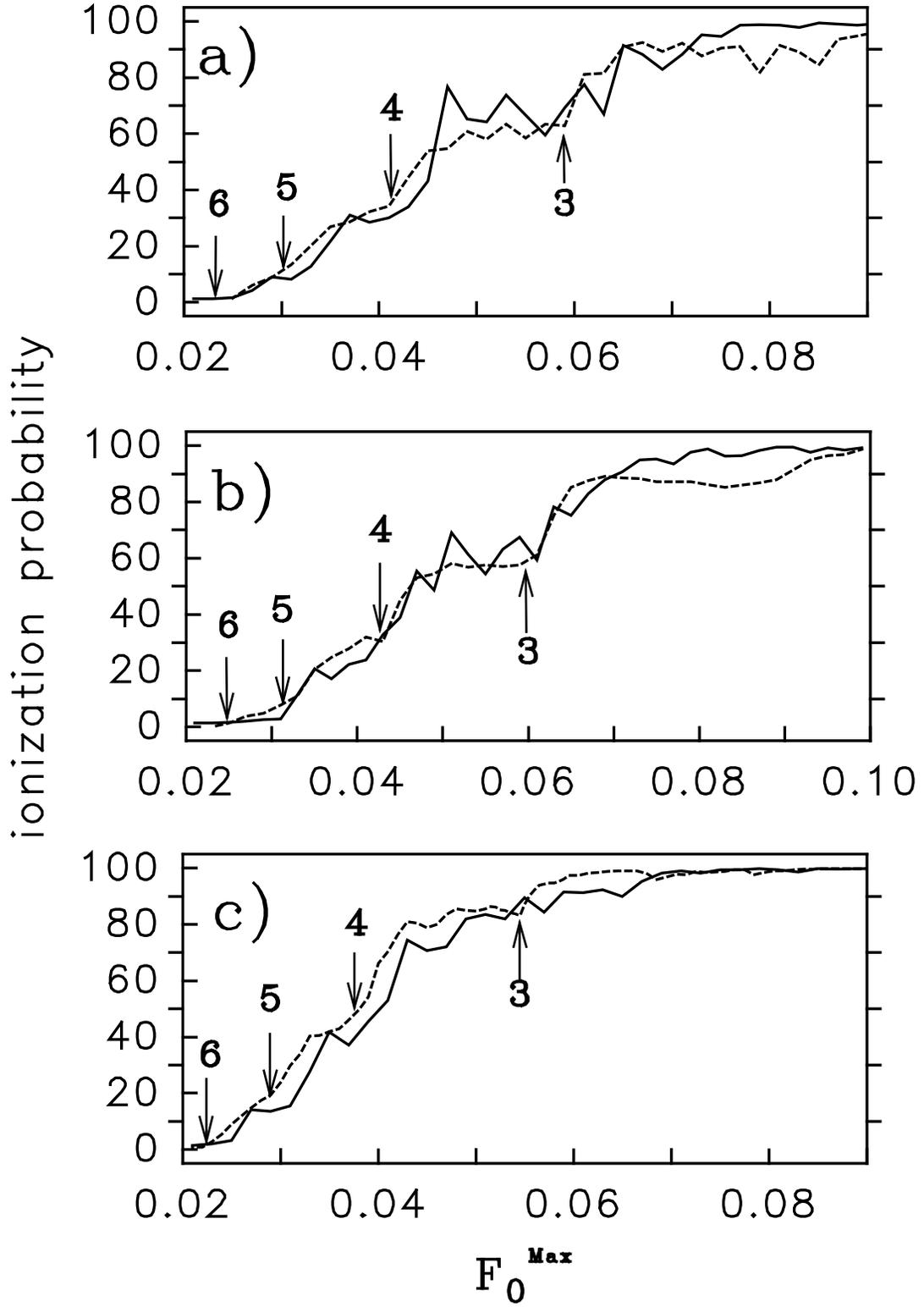,width=0.8\linewidth}
\caption{Classical (dashed lines) and quantum (full lines) steps for
a) $\omega'_0=0.9771$, b) $\omega'_0=0.9994$  and c)
$\omega'_0=0.927$; $F_0^S= 0.035266$. Classical destabilization
fields for secondary island chains are indicated by arrows; they are in all
cases very close to the onset of a ionization step. (From Ref. \cite{ref40})}
\label{fig6}
\end{figure}

In Ref. \cite{ref9}, the large secondary resonance island chains encircling the stable fixed point of the
Poincar\'{e} map at the center of the $\omega'_0 =1$ resonance zone were identified as the cause of the sequence of
steps in the classical ionization probability vs. peak microwave field $F_0^{max}$ graphs, of which three examples
are shown as full lines in Fig. \ref{fig6}. These chains are born from the distruction of those periodic orbits of
the Poincar\'{e} map laying within the $\omega'_0=1$ primary resonance zone which have periods whose ratio to the
microwave period equals the ratio $q/p$ of two (relatively prime) integers \cite{ref9}. As $F_0$ is increased, The
most visible chains have $p=1$ (``integer resonances") and the number $q$ of individual islands in them decreases
with $F_0$ \cite{ref9}. The size of an island chain grows rapidly with $F_0$ above a bifurcation point $F_{0q/p}$
in qualitative agreement with the simple pendulum approximation model described in appendix \ref{app:duea}.

\begin{figure}[htbp]
\centering\epsfig{file=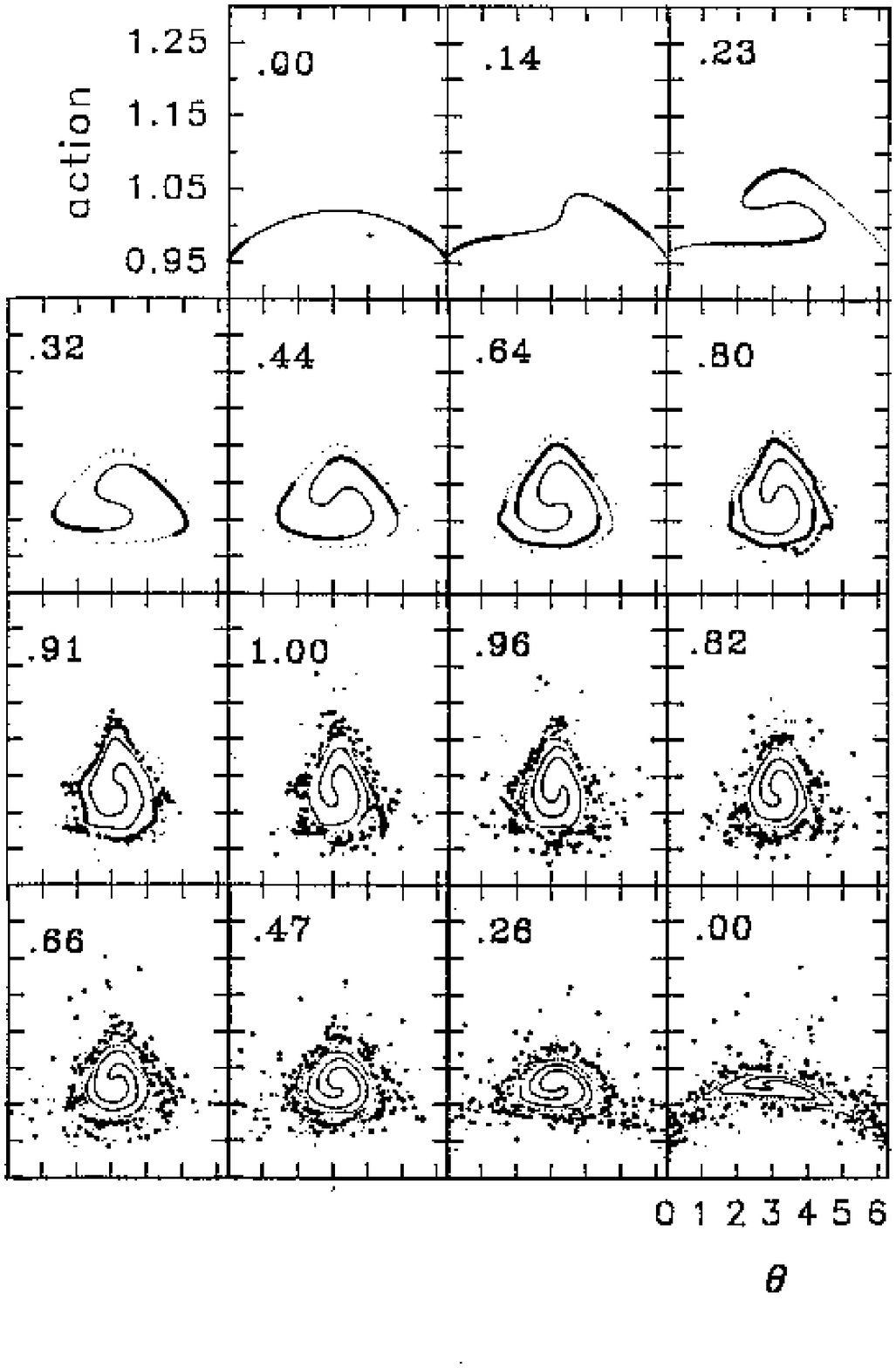,width=0.8\linewidth}
\caption{Snapshots of the evolution of the classical ensemble for the
parameter values $\omega'_0=0.9771$, $F_0^S= 0.035266$ and
$F_0^{max} = 0.025$. The ratio of $F_0(t)$ to $F_0^{max}$ is
indicated in the top left corner of each snapshot. Bigger dots mark
the part of the ensemble within the $q=6$ island chain at the peak
of the pulse. (From Ref. \cite{ref40})}
\label{fig3a}
\end{figure}

\begin{figure}[htbp]
\centering\epsfig{file=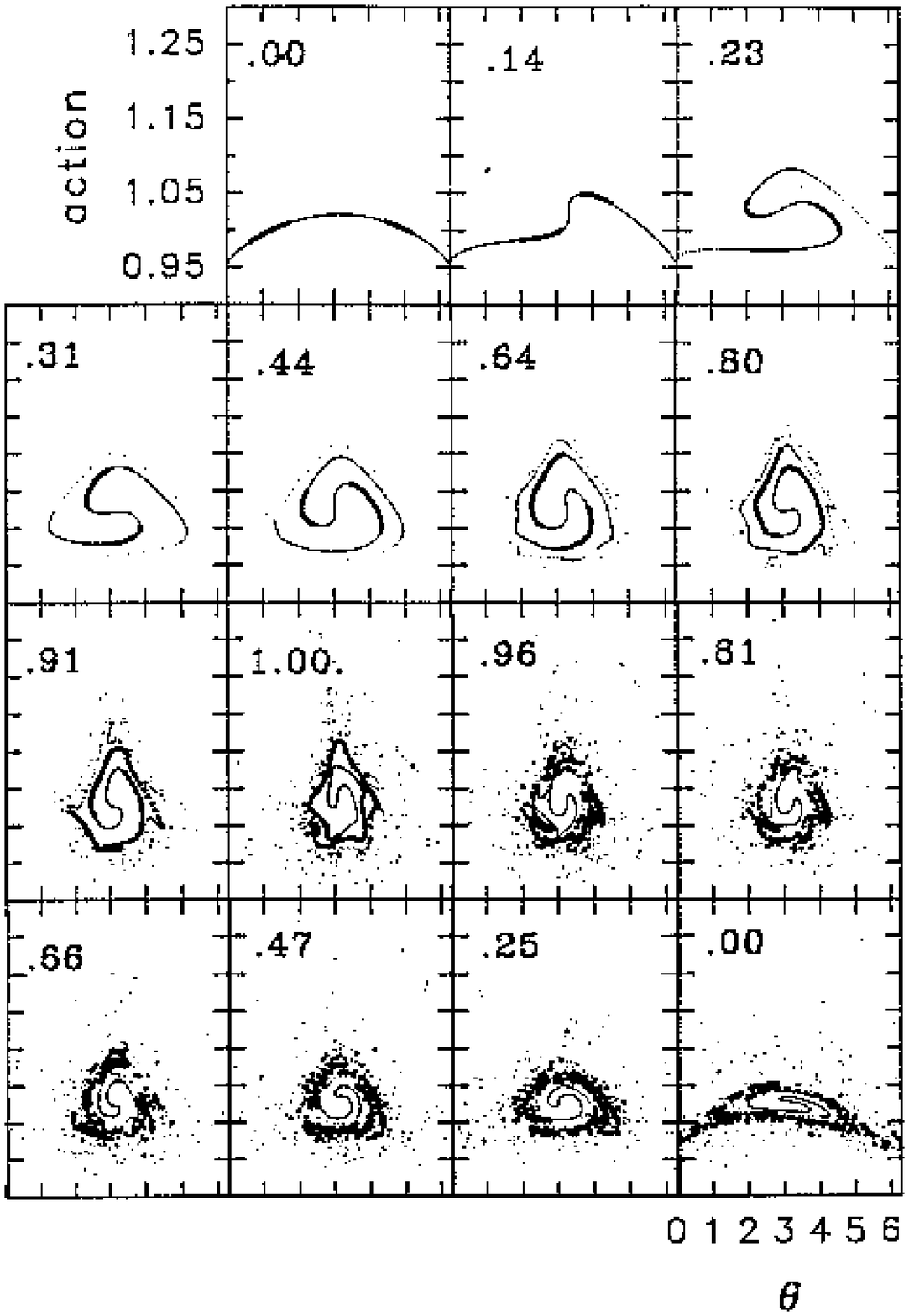,width=0.8\linewidth}
\caption{Snapshots of the evolution of the classical ensemble for the
parameter values $\omega'_0=0.9771$, $F_0^S= 0.035266$ and
$F_0^{max} = 0.029$. The ratio of $F_0(t)$ to $F_0^{max}$ is
indicated in the top left corner of each snapshot. Bigger dots mark
the part of the ensemble within the $q=5$ island chain at the peak
of the pulse. (From Ref. \cite{ref40})}
\label{fig3b}
\end{figure}

\begin{figure}[htbp]
\centering\epsfig{file=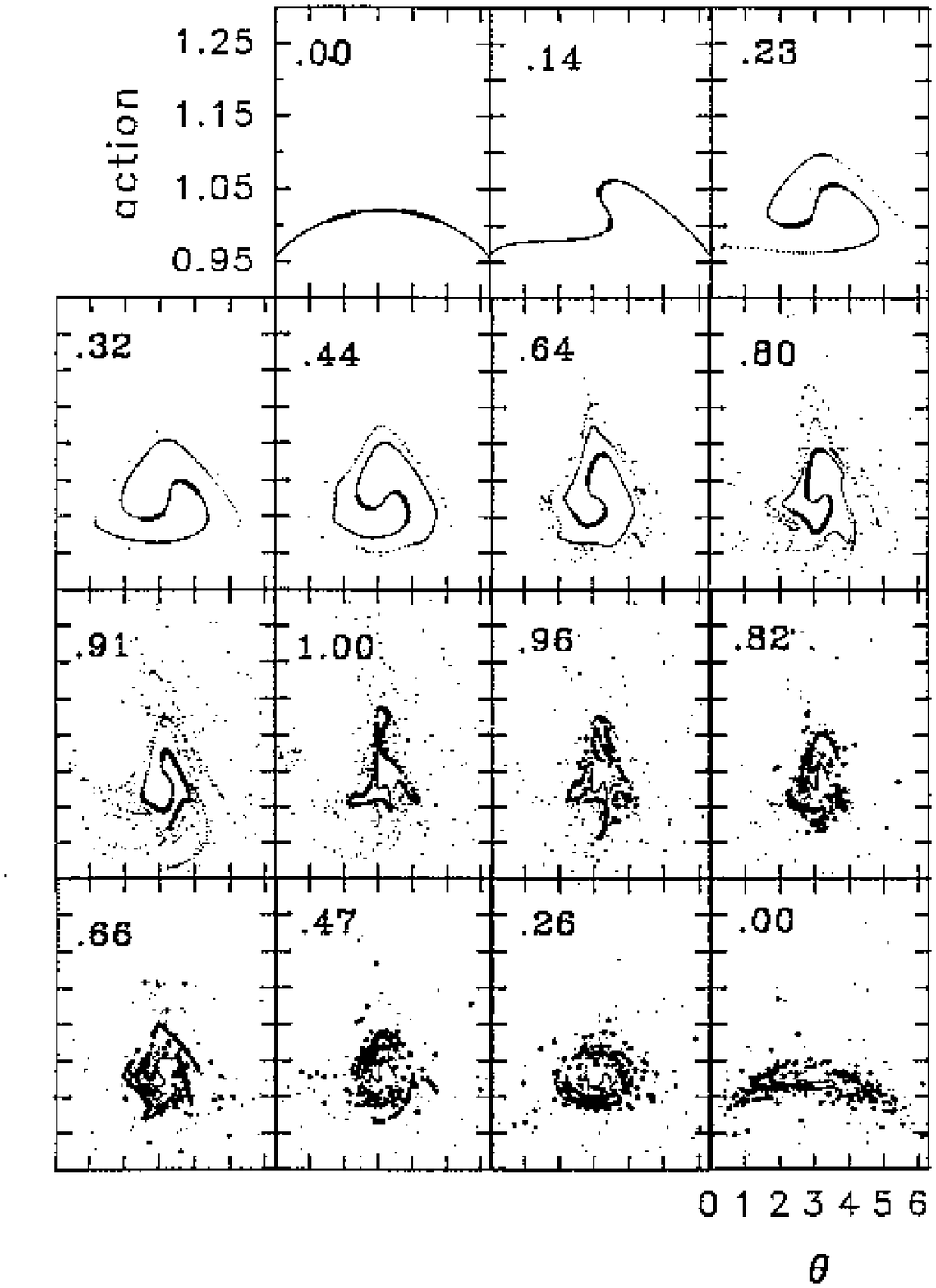,width=0.8\linewidth}
\caption{Snapshots of the evolution of the classical ensemble for the
parameter values $\omega'_0=0.9771$, $F_0^S= 0.035266$ and
$F_0^{max} = 0.041$. The ratio of $F_0(t)$ to $F_0^{max}$ is
indicated in the top left corner of each snapshot. Bigger dots mark
the part of the ensemble within the $q=4$ island chain at the peak
of the pulse. (From Ref. \cite{ref40})}
\label{fig3c}
\end{figure}

\begin{figure}[htbp]
\centering\epsfig{file=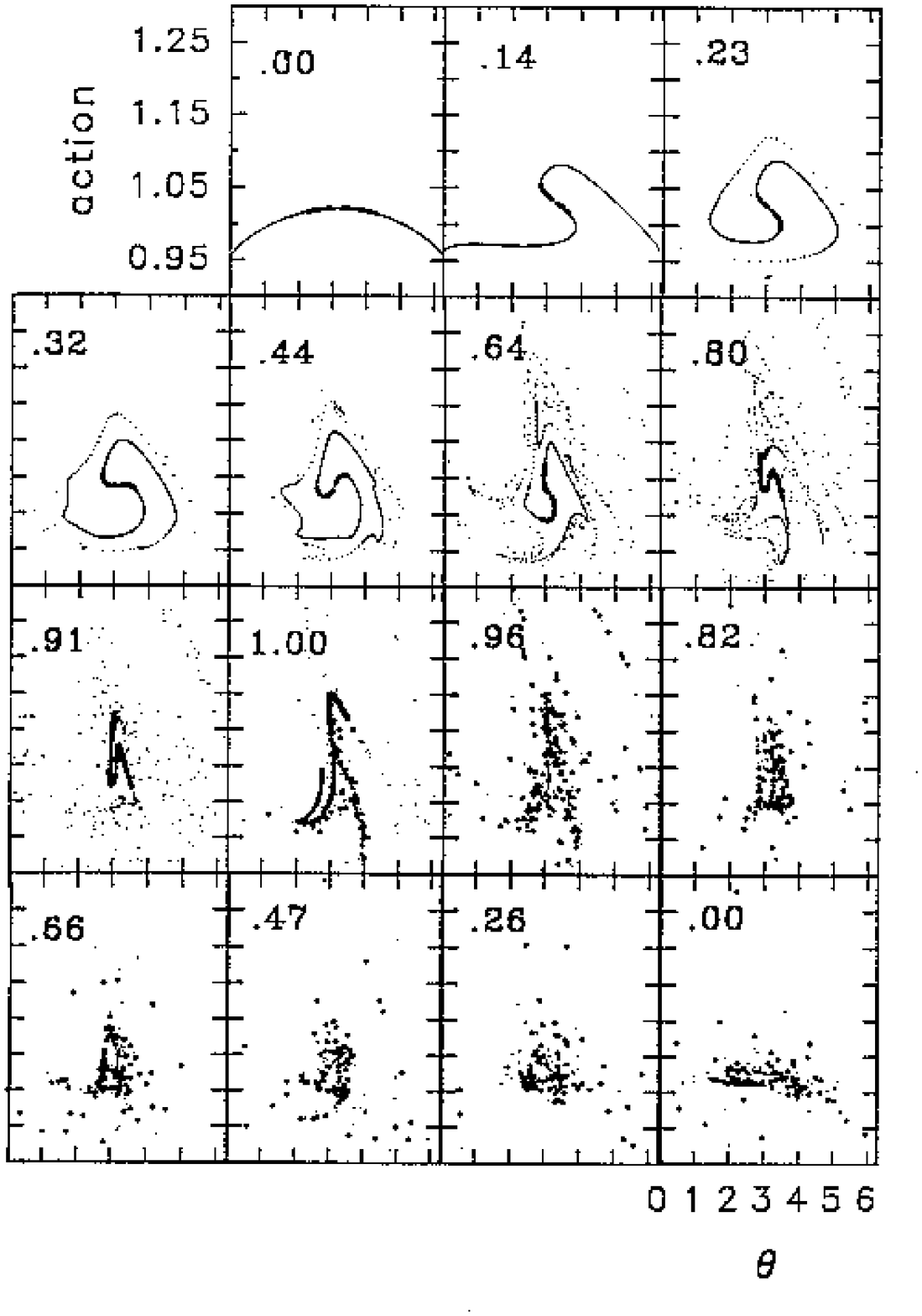,width=0.8\linewidth}
\caption{Snapshots of the evolution of the classical ensemble for the
parameter values $\omega'_0=0.9771$, $F_0^S= 0.035266$ and
$F_0^{max} = 0.061$. The ratio of $F_0(t)$ to $F_0^{max}$ is
indicated in the top left corner of each snapshot. Bigger dots mark
the part of the ensemble within the $q=3$ island chain at the peak
of the pulse. (From Ref. \cite{ref40})}
\label{fig3d}
\end{figure}

\begin{figure}[htbp]
\centering\epsfig{file=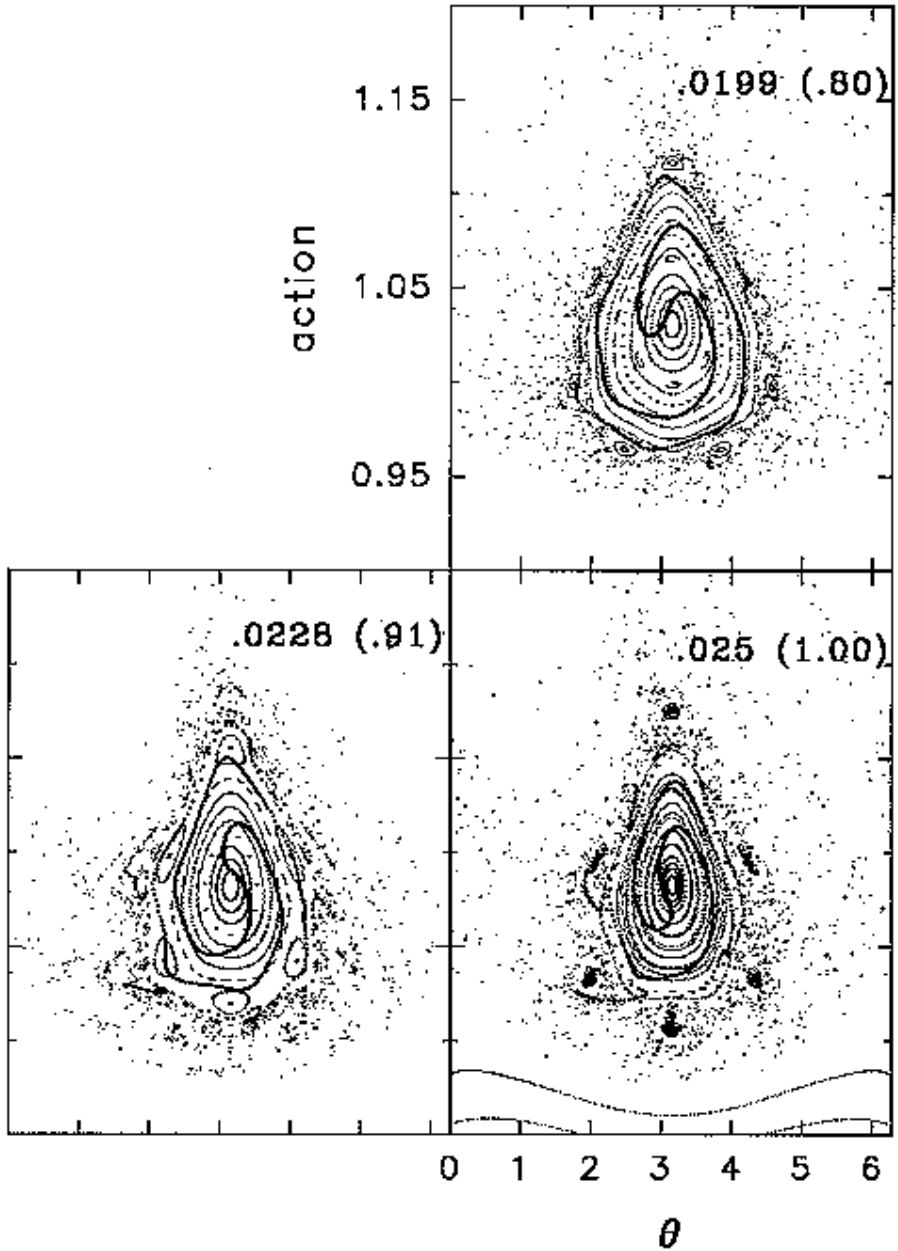,width=0.8\linewidth}
\caption{Comparison of snapshots from Fig. \ref{fig3a} ($F_0^{max} = 0.025$) with the surfaces of section
calculated at the values of the instantaneous $F_0(t)$; $F_0$ is
indicated at the top of each graph, its ratio to $F_0^{max}$ is
given in parenthesis. In the first snapshot three islands of the
$q=8$ chain are deforming the tails of the ensemble; in the second
one there are traces of tendrils of the almost completely destroyed
$q=7$ chain, while the $q=6$ chain is beginning to be felt; in the
last snapshot the $q=6$ chain is almost completely destabilized,
but still some tendrils and (on the left side) the remains of a
whorl, are visible. (From Ref. \cite{ref40})}
\label{fig4a}
\end{figure}

\begin{figure}[htbp]
\centering\epsfig{file=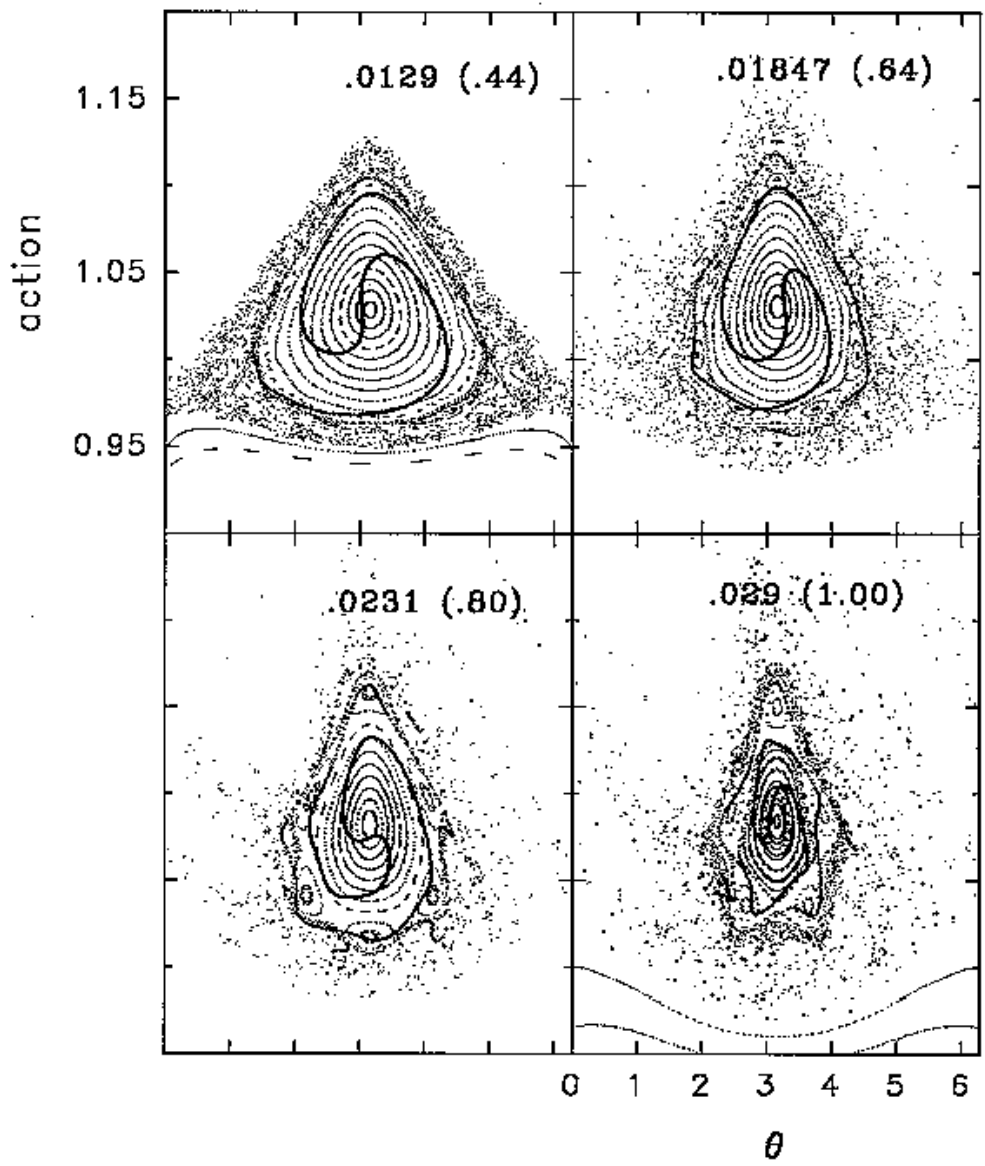,width=0.8\linewidth}
\caption{Comparison of snapshots from Fig. \ref{fig3b} ($F_0^{max} = 0.029$) with the surfaces of section
calculated at the values of the instantaneous $F_0(t)$; $F_0$ is
indicated at the top of each graph, its ratio to $F_0^{max}$ is
given in parenthesis. In the second snapshot the $q=7$ chain is
beginning to be felt; in the third it is the $q=6$ chain that is
deforming the tails of the ensemble; and in the fourth the $q=5$
chain. (From Ref. \cite{ref40})}
\label{fig4b}
\end{figure}

\begin{figure}[htbp]
\centering\epsfig{file=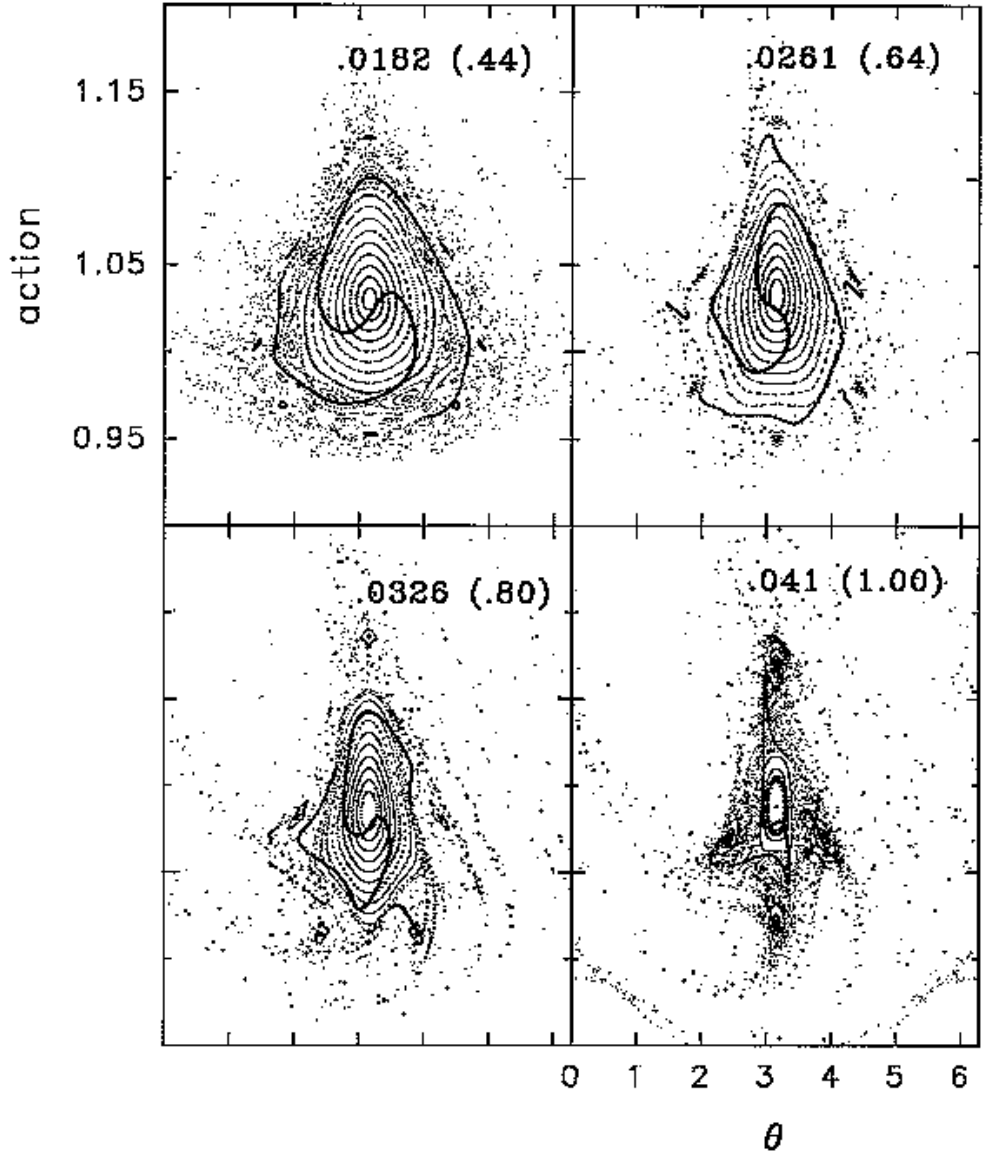,width=0.8\linewidth}
\caption{Comparison of snapshots from Fig. \ref{fig3c} ($F_0^{max}=0.041$) with the surfaces of section
calculated at the values of the instantaneous $F_0(t)$; $F_0$ is
indicated at the top of each graph, its ratio to $F_0^{max}$ is
given in parenthesis. In the first snapshot both the $q=8$ and
$q=7$ chains sligtly act on the enemble; but in the second the
tendrils of the $q=6$ chain are well developed; in the third the
active chain is the $q=5$ one; and in the fourth the $q=4$ one.
(From Ref. \cite{ref40})}
\label{fig4c}
\end{figure}

The arrows in Fig. \ref{fig6} show the field strength values at
which the $p=1$ island chains destabilize (become completely
chaotic) and are destroyed, for $q=3$ to $6$. Each of these values
is very close to a step onset, strongly suggesting a connection
between the ionization steps and the secondary island chains; the
connection is confermed by the evolution in phase space of the
classical trajectory ensemble. Four examples - of the many I have
calculated in Ref. \cite{ref40}- are given in Figs.
\ref{fig3a}-\ref{fig3d}: in all cases we are at the base of an
ionization step and a sizable part of the ensemble (marked by
bigger dots) finds itself at the peak of the pulse inside the
islands of some secondary chain: $q=6$ for Fig. \ref{fig3a}, $q=5$
for Fig. \ref{fig3b}, $q=4$ for Fig. \ref{fig3c} and $q=3$ for Fig.
\ref{fig3d}. Together with the big whorl of the points trapped
within the primary island, the marked points make up most of the
sub-ensemble surviving the microwave pulse, as can be seen from the
last snapshot of each sequence. Whorls and tendrils developed by
the ensemble at intermediate times during the rise of the pulse
indicate the influence of other secondary chains with higher $q$'s.
The relevant chains were in each case identified by comparing the
snapshot of the ensemble with the constant field stroboscopic
surface of section taken at the same value of $F_0$; examples are
shown in Figs. \ref{fig4a}-\ref{fig4d}. These comparisons reveal
the following classical time sequence of events responsible for a
step in Figure \ref{fig6} (a more schematic description was given
in Ref. \cite{ref9}). The ensemble initially is located on a
nonresonant phase-space KAM curve which spans the full range $[0,
2\pi]$ of the angle variable. The pulse peak field $F_0^{max}$ must
of course be high enough that by the time we are at the peak of the
pulse the locally chaotic region of the principal primary resonance
has merged with the global chaos above (the last KAM curve
separating them has broken) so that classical ionization is
possible.

A) For $F_0(t)$ near $0.01$, the ensemble almost entirely crosses
the locally chaotic region to enter the $\omega'_0=1$ island
region; the wide ``nose" the ensemble develops at the crossing (see
the second snapshots in Figures \ref{fig3a}-\ref{fig3d}) tells us
that after the crossing the spread in resonance action $J$ (action
of the slow motion around the stable fixed point of the
Poincar\'{e} map) is quite wide.

B) After the crossing there is a time interval of slow near
adiabatic ensemble evolution, having an approximate frequency given
by eq. (\ref{eqb9a}). The ensemble is at the same time partially
``adiabatically squeezed" in the angle variable \cite{ref24} toward
the center of the island (fourth to sixth snapshots in Figures
\ref{fig3a} and \ref{fig3b}, third to fifth  snapshots in Figure
\ref{fig3c} and third to fourth snapshots in Figure \ref{fig3d}).
If $|\omega'_0-1|<0.05$, much of the ensemble approaches and stays
close to the center of the island.

C) A bifurcation sequence with decreasing $q$ has been occurring
with increasing $F_0(t)$. For each $q$ the size of the
corresponding island chain grows rapidly \cite{ref9}. Within a
field strength increment between $5\%$ and $20\%$, this size
reaches a value about equal to the nominal size of the entire
primary island region  ($\Delta n=8$ to $11$ depending on
$F_0(t)$). The size of the individual islands of the chain grows
too; for a given size $\Delta I$ of the chain the size of its
islands appears to increase with decreasing $q$.

D) A study of surfaces of section at $F_0 \lesssim 0.01$ (crossing
field strength) shows that the secondary island chains with $q
\gtrsim 8$ are all born before the crossing; most of them (those
with $q \gtrsim 12$) have moreover already grown to full size and
destabilized. The remaining secondary island chains can and - for
$q<8$ - do influence the evolution of the ensemble as can be seen
in the superpositions of ensembles over instantaneous surfaces of
section shown in Figs. \ref{fig4a}-\ref{fig4d}: parts of the
ensemble are ``trapped" by these chains. By ``trapped" I mean that
once the points of a part of the ensemble are inside the islands of
a chain, they remain inside those islands and follow the chain in
its motion (during the rise of the pulse) toward the periphery of
the primary resonance zone \cite{ref46}. Note that the direction of
this motion is the opposite of that of the adiabatic ``squeezing"
we observed in step B. This inversion of motion of parts of the
ensemble within the primary resonance island is a clear indication
of their trapping by the secondary resonance zone. Finally, since
the span, in the resonance action $J$, of the islands of the
secondary chain is usually not big, it is important that a sizable
portion of the ensemble be at some time within the same small range
of $J$ so as to be able to be trapped by the islands of the
secondary chain. This is made possible by the  ``squeezing" we have
seen at point B.

For each of the four cases in Figs. \ref{fig3a}-\ref{fig3d} I have
marked with darker dots that part of the ensemble which at the peak
of the pulse is within the relevant island chain: the motion of the
``trapped" part of the ensemble toward the periphery of the primary
resonance zone is particularly clear in Figures \ref{fig3c} and
\ref{fig3d}. It is also evident that different island chains act on
different parts of the ensemble: the higher is the $q$ of a chain
the further from the center of the primary resonance zone is the
part of the ensemble it ``traps". Parts of the ensemble closer to
the center of the resonance zone do not ``see" the chain when it
grows through them: they are traversed diabatically (motion within
the secondary islands is too slow to be ``seen" by the points of
the ensemble for the chosen switch-on time). Only at a certain
distance from the center of the primary resonance zone a given
island chain becomes ``visible" to the ensemble, and this leads us
to believe that, together with the size of the chain and the size
of the islands of the chain, also the (average) frequency of motion
within those islands must increase with $F_0$ (just like the
frequency of motion within the primary island: see eq.
(\ref{eqb8b})). Moreover, for a given pulse length, the distance
from the center at which this happens increases with the $q$ of the
chain, suggesting that, for any given distance from the center, the
higher is $q$ the lower is the frequency of motion on the islands
of the chain.

Now that we have part of the ensemble trapped in the secondary
chain, we have two possibilities : E and E'+F'.

E) As long as $F_0^{max}$ remains smaller than the one necessary to destroy
the islands of the secondary chain, the only part of the ensemble which can
and does ionize is the one in the chaotic region of the primary resonance
zone; no further ionization is possible. We are on the plateau of the step.

E') If we instead further increase $F_0^{max}$, the island chain
becomes locally chaotic, and then ``globally" chaotic (that is: it
merges with the chaotic band surrounding the primary resonance
island and finally with the global chaos above), all within a
further field strength increment of about $5\%$: rather abruptly a
sizable part of the phase space becomes globally chaotic. It is
this sudden increase of the globally chaotic region that causes the
step in ionization probability. The growth of the band of local
chaos is reflected in the changed appearance of the parts of the
ensemble trapped in the chain: whorls, connected to motion within
the stability islands, are substituted by tendrils, connected to
motion in the chaotic band. the last two snapshots in Fig.
\ref{fig4a} show such a transition for the $q=6$ island chain; the
association of tendrils with the destruction of chains is also
clear in the second snapshot in Fig. \ref{fig4c} (again the chain
is the $q=6$ one) and in the third ($q=5$), fourth ($q=4$) and
sixth ($q=3$) snapshot in Fig. \ref{fig4d}.

F') The chaotic trajectory subensemble released by the island chain
destruction event E' produces ionization during a  time interval of
about $30\%$ of the pulse length, within the relatively constant
central part of the half sinewave pulse.

The examples I have given in Figures \ref{fig3a}-\ref{fig3d} and
\ref{fig4a}-\ref{fig4d} make clear that ``trapping" in the
secondary island chains is possible for the experimental pulse
times of about $131$ microwave periods. Longer pulse times mean
slower growth of $F_0(t)$ resulting in even more effective
trapping, as can be seen from Fig. 14 in Ref. \cite{ref25}: the
sharpness of the ionization probability steps improves with
increasing pulse times.

An important implication of the above scenario is that, for fixed
values of the static field $F_r^s$ (rescaled at the resonant action
$I_r$ defined by $\omega I_r^3/(1 - 3F_s I_r^4) = 1$), the position
in $F_r^{max}$ of the steps depends only on the destabilization of
the island chain and not on the initial conditions of the ensemble.
The initial action (and the pulse time) on the other hand determine
the heights of the steps. As an example I have plotted classical
ionization probability data for rescaled frequencies $\omega'_0$
between $0.9325$ and $1.0666$ first as a function of $F_0^{max}$
(Fig. \ref{fig5}.a) and then as a function of $F_r^{max}$ (Fig.
\ref{fig5}.b): notwithstanding the great difference in height, the
position of the steps in Fig. \ref{fig5}.b is approximately the
same ($F_r^s$ is slightly different from curve to curve); while
this is clearly not the case in Fig. \ref{fig5}.a \cite{ref26}.

Parallel quantum numerical simulations are shown as dashed curves
in Fig. \ref{fig6}. In Figure \ref{fig6}.a. The classical steps
labeled $q=3-6$ are reproduced quite well. They appear even sharper
in the quantum simulations than they are in the classical ones. On
the other hand the oscillations of the plateau just before the
$q=3$ step are much more pronounced than the classical ones. The
absence of a $q=2$ quantum step in the figure most likely reflects
the fact that the $q=2$ island chain in (sectioned) phase space is
not large enough to significantly support a Floquet eigenstate
(even if their height is comparable to that of the islands of other
chains, the two islands are extremely narrow: never more than about
$0.1\hspace{.1in} radians$ in the angle variable). Figure
\ref{fig6}.b shows another case where the above comments again
apply: the quantum curve follows quite closely the classical one up
to the $q=3$ step but then directly rises to $100\%$ ionization
probability, skipping the $q=2$ step. Figure \ref{fig6}.c finally
shows the quantum and classical steps for a third case; this time
$|\omega'_0-1|= 0.073$ is big enough that the ensemble does not
penetrate all the way to the island's center. I believe this to be
the cause of the absence of the $q=2$ step in the classical curve:
by the time the $q=2$ island chain is born, the primary island is
so small that all of the ensemble is outside of it in the chaotic
region and cannot be trapped.

The good agreement between classical and quantum simulations is not limited
to the final ionization probability; the evolutions during the pulse are
remarkably similar too. In Fig. \ref{fig7} I show a comparison of the quantum and
classical ionization probability versus time: the shape of the curves is
always very similar even if the final ionization probability is often not
exactly the same.

\subsubsection{Classical and quantum resonances}

To explain this agreement between classical and quantum ionization
results, we now explore the connection between higher order
classical resonances and quantum resonances at nonzero microwave
field (avoided crossings in the quasienergy plots). We start with
some general considerations which will then allow us to locate the
avoided crossings related to the secondary resonance zones within
the principal primary resonance island \cite{ref40}.

For simplicity, let us start considering an isolated avoided
crossing of two {\bf energy} levels $E(n_i,F)$, $i=1,2$ with $F$
some perturbation parameter. We assume that we have been able to
separate the complete Hamiltonian $H$ of the system in two parts,
$H_0$ and $H_1$, such that only the interaction responsible for the
separation of the levels at the crossing is included in $H_1$ and
the levels $E_0(n_i,F)$, eigenvalues of $H_0$, actually cross at
the field $F_c$ of closest approach of the levels $E(n_i,F)$ of
$H$. We moreover assume that $E_0(n_i,F)$ is the quantized version
of a classical function $E_0(I,F)$ which is smooth for $I \in
[n_1,n_2]$. Since the classical phase space can be divided into
regions corresponding to different kinds of motion  and
discontinuities in action occur at the borders of these regions
\cite{ref27}, the above condition requires us to assume that for
all relevant $F$'s the two levels of $H_0$ we are considering
correspond to the same kind of classical motion (their support is
in the same region of the classical phase space of $H_0$). When
these conditions are verified, at the field strength value $F_c$
where the levels $E_0(n_i,F)$ cross the average slope of the
function $E_0(I, F_c)$ for $I\in [n_1, n_2]$ is $0$. There must
therefore exist an action $I_r \in [n_1, n_2]$ verifying the
classical resonance condition $\partial E_0(I,F_c)/
\partial I|_{I_r}=0$ (Rolle's theorem). For such an avoided crossing and for small
values of the perturbation, a relationship has been derived in Ref.
\cite{ref28} expressing the splitting at the avoided crossing as a
function of the area of the related resonance zone, both functions
of the interaction Hamiltonian $H_1$.

If the perturbation Hamiltonian $H_1$ is time dependent and
periodic with frequency $\omega$, then the avoided crossing is
between quasienergy states, the quantum resonance condition for the
(no more crossing) energy states of $H_0$ is eq. (\ref{eq7}) and the
classical resonance condition is $\partial E_0(I,F_c)/ \partial
I|_{I_r}= s\omega/r$.

Let us now take $H_0$ to be the ``pendulum approximation" Hamiltonian
\begin{equation}
H^{(1)} =  [- {\frac 1 {2 I^2}} - {\frac 3 2}F_sI^2  - \omega I] +
0.325FI_r^2 \cos{\varphi}\nonumber
\end{equation}
which describes the resonant motion within the principal primary
resonance zone $s/r=1$ (see appendix \ref{app:dueb}). Its
numerically calculated levels for the parameter values of Fig.
\ref{fig6}.a are shown in Fig. \ref{fig8} (again see appendix
\ref{app:dueb} for details). After each level's inflection point
(which is when the level gets ``trapped" \cite{ref20} into the
region of libration motion), the functional dependence of the
energy of the eigenstates of $H^{(1)}$ on the pendulum quantum
number $k$ is that of the classical energy on the action $J$ within
the (primary) ``resonance" zone of libration motion (see fig
\ref{figb1} in appendix \ref{app:dueb}). Since this dependence is
smooth we can apply the argument above: crossings between states
with $\Delta k=q$  and energy distance $\Delta E=p \omega$ are
related to the $\Omega =
\omega p/q$ classical secondary resonance condition where $\Omega$
is the ``slow motion frequency" $|\partial H/ \partial J|$. In Fig.
\ref{fig8} I have marked the crossings of the pendulum
approximation levels with different symbols for the different $p/q$
ratios (connected by lines to guide the eye); arrows mark the
classical bifurcation fields eq. (\ref{eqb9b}) for the same $p/q$
ratios. As expected from the argument above, all the crossings with
a given $p/q$ ratio lay to the right of the (pendulum
approximation) bifurcation field for the secondary resonance zone
with the same $p/q$ ratio. Also, the crossings with the same $p/q$
ratio are not all at the same field but the higher the average $k$
of the levels crossing the higher the field at which the crossing
takes place, in agreement with the moving (with increasing $F$) of
the center of the classical secondary resonance zone toward higher
pendulum actions \cite{ref40}.

We can now turn to the quasienergy levels of the full problem, of
which the pendulum ones are a good approximation near the $s/r=1$
quantum nonlinear resonance at low microwave field strengths
\cite{ref29}. The true levels' avoided crossings corresponding to
the approximate levels' crossings, even if shifted (at high
microwave field strengths) with respect to these latter ones, are
related to the same secondary resonance zones and therefore lay to
the right of the numerical classical bifurcation fields for the
same $p/q$ ratios, as can be seen in Fig. \ref{fig1} (where we use
the same conventions as in Fig. \ref{fig8}) \cite{ref40}.

The identification of the crossings related to these secondary
island chains is supported by the Husimi functions I calculated for
the Floquet eigenstates undergoing the avoided crossings: some of
them are shown in Figs. \ref{fig9a}-\ref{fig9e}. Before discussing
them let us first introduce one last concept: that of {\it crossing
region}. For $F\ll F_c$ the two quantum states $E(n_i,F)$ are
practically noninteracting and coincide with $E_0(n_i, F)$;
approaching $F_c$ they begin to interact more. This interaction is
reflected in the mixing of the two states which is maximum at $F =
F_c$  where the two eigenstates of $H$ undergoing the avoided
crossing are linear combinations in equal parts of the two
(crossing) eigenstates of $H_0$. After that, the mixing decreases
until, for $F\gg F_c$, they are again the two pure $E_0(n_i, F)$
states we started with but now the state $n_1$  has the spatial
structure $n_2$ had before the crossing and viceversa, as if the
two levels had switched. The quantum resonance itself as defined by
eq. (\ref{eq7}) is therefore only the central point of a region in
$F$ where the two levels interact with each other. The values of
$F$ such that the separation of the noninteracting levels
$E_0(n_i,F)$ equals the interaction term can be taken as
approximate boundaries in $F$ of the region where there is
significant mixing \cite{ref19}; we call this region the crossing
region. Away from crossing regions the Husimi functions of the
primary resonance eigenstates appear approximately as circular
ridges, apart from the $k=0$ one which is approximately a Gaussian
peak \cite{ref19}. All these ridges have their center in the center
of the classical principal primary island and the higher the
resonance quantum number $k$, the higher their average radius.
Examples are the Husimi functions before and after the avoided
crossings shown in Figures \ref{fig9a}-\ref{fig9e}. At a crossing
related to classical secondary chain of $q$ islands, the exchange
of the spatial structures of the two eigenstates involved happens
in a characteristic way: the outer ridge moves inwards avoiding
(``seeping" around) the $q$ regions of the islands of the chain
while the inner ridge moves outwards right through these regions,
avoiding the regions around the unstable periodic orbit of the
island chain and sometimes developing $q$ peaks corresponding to
the islands of the secondary chain. Often this $q$-fold structure
of the states is not equally visible for both states at the same
value of the microwave field strength $F$: for most of the cases
studied the structure is most visible for the outer states at
slightly lower values of $F$ than for the inner state. Fig.
\ref{fig9a} is a particularly clear case: the Husimi functions of
the $k_1=2$ and $k_2=7$ resonance states both develop at their
crossing the expected fivefold structure. Figure \ref{fig9b} shows
instead the ($q=7$) crossing of the $k_1=3$ and $k_2=9$ states:
since the $k_2=9$ state is here almost a separatrix state (its
Husimi function has not yet taken the typical resonance state shape
and is peaked around the unstable fixed point of the Poincar\'{e}
map \cite{ref23}) we are nominally outside the range of
applicability of my semiclassical argument, still the sevenfold
structure is visible. In other cases the process can be complicated
by the interaction with other states but the $q$-fold structure is
always present. Figure \ref{fig9c} shows the ($q=6$) crossing of
the states $k_1=2$ and $k_2=8$ where the lower ``adiabatic" state
(with respect to the $q=6$ crossing) is made up of the two
interacting states B and C. In Fig. \ref{fig9d} ($q=5, k_1=0,
k_2=5$) it is the upper ``adiabatic" state which instead is made up
of the two states A and C. Finally Fig. \ref{fig9e} shows a case
($q=4, k_1=0, k_2=4$) where the three level interaction results in
the adiabatic state ``B" assuming before the crossing the role of
upper ``adiabatic" state and after the crossing the role of lower
``adiabatic" state, thus maintaining the same resonance
quantization $k_2=4$. These Husimi functions make clear that at the
crossing regions I have marked in Fig. \ref{fig1} the crossing
eigenfunctions do see (and reflect in their structure) the $q$ dips
of the classical dynamical potential \cite{ref30} evidentiated in
the surfaces of section by the $q$ islands of the secondary chain
they are related to.

We now have to show that these crossings are in effect those where
most of the population transfer causing the ionization steps
happens during the microwave pulse. To do this we plot the
projections of the wavefunction on the instantaneous Floquet
eigenstates at each period of the microwave during the pulse. For
the four values of the peak microwave field strengths of Figs.
\ref{fig3a}-\ref{fig3d} these projections are shown in Figs.
\ref{fig10a}-\ref{fig10d}. They give us a dramatic representation
of the importance of the avoided crossings we have individuated
(marked here with the same symbols as in Fig. \ref{fig1}). After
the initial spread of the population over the states $k=0-3$, the
population is further spread out to high $k$ short lived states
through these avoided crossings or sometimes through avoided
crossings which can be related to secondary island chains with
higher $q$ than those marked. So in Fig. \ref{fig10a} the two
(couples of) crossings which are not marked are those between $k=1$
and $k=9$ and between $k=2$ and $k=10$; both can be reconducted to
the $q=8$ secondary island chain but only by analogy with the cases
studied, since the crossings happen early in the $k=9$ and $k=10$
curves and their WKB quantization is not yet the resonance one
(i.e. depending on the quantum number $k$). The ($q=8$) crossing
between $k=2$ and $k=10$ appears again as significant in Fig.
\ref{fig10b} but in Fig. \ref{fig10c} the transfer of probability
at that crossing has dwindled to almost nothing and the only
significant crossings are the marked ones. Finally in Fig.
\ref{fig10d}, corresponding to a case where the classical principal
primary resonance zone is almost completely chaotic at the peak of
the pulse, the marked crossings are again the first relevant
crossings and the only ones to cause significant transfer of
probability; but several other crossings cause some (minor)
transfer of probability at the peak of the pulse. The most
important of these latter crossings are related to classical
secondary ``fractionary" resonances: for example the crossings
marked by crosses (from top to bottom, between $k=0$ and $k=9$,
between $k=1$ and $k=10$ and between $k=1$ and $k=11$) have $p=2$.
There also appears to be some overlap of significant crossings each
involving the same level but related to different classical
resonance zones (for example the crossings between $k=1$ and $k=5$
and between $k=1$ and $k=11$ in the bottom right corner of the
figure); this is by some authors thought to be a sign of an
analogous overlap of the related classical resonances and therefore
an expression in quantum dynamics of local classical chaos
\cite{ref31}. To emphasize the step-like character of the evolution
of the $k=0-3$ states, I give in Table \ref{table1} their
population loss at each of the relevant crossings.

\section{Discussion}

We have thus shown the relationship between some strictly quantum processes (avoided crossings)
and classical structures; but why do isolated avoided crossings so well simulate the classical
behaviour in the cases we studied? To understand this we should try and get an idea of the
circumstances under which the classical and quantum dynamics can be expected to result in
similar evolutions of the observables.

Let us start considering the system at a {\it given microwave field
strength}; when can we say that the properties of the quantum
system reflect the classical resonance structure? Away from any
crossing region related to a given secondary classical resonance,
the quantum system is by definition not affected by that classical
resonance zone. Only for values of $F$ falling within one such
crossing region  the quantum eigenstates reflect the classical
resonance region, as we have seen.  If the crossing regions related
to the classical resonance zone are well spaced, it will therefore
be only by a chance choice of $F$ that the quantum dynamics will be
able to ``feel" the classical resonance zone. Only when the
crossing regions related to the classical resonance zone overlap
this element of chance is lost.

If we look at Figure \ref{fig1} we see that there is hardly any
overlap of the crossing regions for any of the $p/q$ ratios marked
in the figure; unless $F$ is carefully chosen we therefore expect
the fixed microwave field strength quantum dynamics not to be
affected by the classical secondary resonance zones for our choice
of $n_0$; in other terms: we cannot be sure of having any quantum
state reflecting (mimicking) the classical secondary resonance
structures. This is essentially \cite{ref32} the base of the
objections which have been raised to a classical interpretation of
the steps \cite{ref25}, notwithstanding the good agreement of
classical and quantum ionization probabilities. I disagree on the
ground that arguments for an atom interacting with a constant
amplitude microwave field are of limited relevance to the present
{\it pulsed} system: since the microwave field strength $F(t)$
changes during the pulse, the system in general meets not just one
but many avoided crossings related to a single classical resonance
(see the examples in Figs. \ref{fig10a}-\ref{fig10d}). Most of
these crossings are between completely different states (e.g., for
the $q=6$ classical resonance, between the states $k=0$ and  $k=6$
, between the $k=1$ and the $k=7$ ones and so on ...) and each of
these avoided crossings is a potentially very quantum event as we
do know that, when a crossing is met by the system twice (once
during the rise and the second time during the fall of the pulse),
the different phases accumulated by the two parts of the
wavefunction (which split at the first crossing) cause interference
at the second crossing, resulting in Stueckelberg-like oscillations
in the population of the quasienergy states as a function of
$F^{max}$ \cite{ref33}. We do not see any trace of this quantum
behaviour. I believe the cause of this to be twofold. First of all,
for the current value of the pulse length $T$, the separatrix
crossing event early in the pulse almost immediately significantly
spreads out the population on states with resonance quantum number
up to $k=3$ as can be seen in Figures \ref{fig10a}-\ref{fig10d}
\cite{ref47}. This means that, even if there were Stueckelberg
oscillations, the ionization signal vs. $F^{max}$ would be the sum
of oscillations due to the couples of crossings met by all four
these states, each of the four states meeting a sequence of them.
The period of each of these oscillations depends on the phase
difference accumulated by the two crossing states between the
crossings, the amplitude on the crossing width (and the speed of
the crossing)  and the phase on the $F^{max} $ when the crossing is
first met by the system. That these oscillations should sum up to a
coherent oscillation, rather than average out  would be a rare
event indeed. Were this the case in our simulations, we should be
able to see some sign of these oscillations in the populations of
the single states. Namely we would expect to sometimes see at the
avoided crossings in the falling edge of the pulse a return of
probability from the high $k$ states to the  four initially
populated ones ($k=0,1,2$ and $3$). While this transfer is visible
at the very end of the pulse among the four states themselves, we
have no evidence of this happening at the avoided crossings we are
considering, as can be seen from Table \ref{table1}. A posible
cause (our second reason for quasi-classical behaviour) is the very
low population on the high $k$ states at crossings on the falling
edge of the pulse: at low peak microwave field strengths only high
$q$ (narrow) avoided crossings are met so that the population
transfer on the rising edge of the pulse is very small. With
increasing peak microwave field strength wider crossings,
corresponding to lower $q$'s, are met and population transfer on
the rising edge becomes substantial, but so does also the decay
rate of the high $k$ states between the two crossings. The
importance of the spread of the initial population can on the other
hand be inferred from the comparison made in Ref. \cite{ref25}
between the classical and quantum final ionization probabilities
for the two pulse times $T=131$ (our case) and $T=300$ microwave
periods. In both cases the agreement increases with increasing
principal quantum number $n_0$; but a good agreement is reached at
lower values of $n_0$ for $T=131$. This latter is the case where we
expect more primary resonance quantum states to be excited at the
crossing of the principal primary island separatrix. In this
picture, rather than being traces of Stueckelberg oscillations, the
fluctuations of the plateaus of the quantum steps in Fig.
\ref{fig6} are likely due to the variations of population transfer
at the avoided crossings during the rising edge of the pulse. For
any given $q$, the first crossing to be met involves the $k=0$
state and, when first met, happens at the peak of the pulse, so
that the interaction time is long and the population transfer high
\cite{ref33}: we have a ionization peak. With increasing peak
microwave field strength the interaction time at the crossing
decreases and so does the population transfer; but then the $k=1$
state has a crossing with the same $q$ and a new ionization peak
appears on the tail of the $k=0$ one, and so on.

\section{Conclusions}

My quantum simulations show that each of the experimentally
observed ionization steps is due to a group of avoided crossings
between couples of Floquet states belonging to the same quantum
nonlinear resonance. These crossings are in their turn related to
the classical secondary resonance island chains surrounding the
principal primary resonance, as the quantum and classical resonance
conditions are formally the same. Finally, the above island chains
are responsible for the steps observed in the classical
simulations, as their distruction releases into the globally
chaotic phase space region (and therefore to eventual ionization)
those parts of the electron ensemble previously trapped within
them.

The good agreement between quantum and classical simulations is a
consequence of the spread - early during the pulse - of the quantum
probability to several resonant Floquet states. This spread allows
more than one of the avoided crossings related to each of the
secondary island chains to redistribute the electron wavefunction
to short lived states, thus contributing to the ionization step
corresponding to that island chain.

My study allows us to conclude that classical phase space
structures too small to influence (in most cases) the quantum
evolution for a constant amplitude microwave field, can -in a
pulsed regime- generate interesting quantum dynamics. Constant
amplitude arguments to determine their relevance should therefore
be used with extreme care, as they can lead to wrong conclusions.

\section{Acknowledgements}

The author wishes to thank C. Rovelli, G. Mantica and S. Locklin for useful comments and
suggestions and the latter also for the use of his computer for the quantum numerical
simulations presented in this paper. Special thanks to J.E. Bayfield as advisor of my Ph. D. thesis, amply quoted in the present paper.

\appendix
\section{}
\label{app:uno}

I discuss here the choice of integration parameters for my
quantum simulations \cite{ref40}. Let us start with considerations about the extremes of the truncated
basis I used. An occupation probability (at every time during the
simulation) of the order of $10^{-10}$ on the lowest two or three
states can be considered a sign that the basis lowest level
$n_{min}$ is small enough. For all of my runs with $n_0$ in the
range $60$ to $72$, $n_{min}= 30$ satisfied this condition.

For the upper bound the matter is more delicate. The spectrum of
atom in the static field alone consists of resonances (bands) whose
number is not well determined, as increasingly wide ones exist with
energies extending above the classical static field barrier. Even
if the free atom states $\psi_n$ of which my basis is composed
extend to infinity their probability distribution is centered
around $\left<z\right>_n=3n^2/2$ with a spread $\Delta z_n\cong
n^2/2$. This means that truncating the basis puts us in a ``fuzzy
box". The potential is therefore a kind of ``double well" where not
only the width but also the depth of the second well varies with
the number of levels considered and the calculated spectrum shows
couples of quasi degenerate double well states instead of bands.
The narrow separation of the couples of levels - up to just below
the classical ionization theshold $E_{max} = -2 F_s^{1/2}$ - tells
us that in the static field alone there is very little coupling
between the two sides of the well (as expected for an asymmetric
double well \cite{ref30,ref34}). In my pulsed simulations, I take
into account the loss of probability induced by the static field by
the substitution in equation (\ref{eq4}) of $E_n$ with
$(E_n-i\Gamma_n/2)$, where the decay factors $\Gamma_n$ are given
by the (3D) formula from Ref. \cite{ref35}; this is a rather crude
approximation: the decay factors are for the Stark states of the
problem and I apply them to free atom states but still using Stark
perturbative energies in calculating them. This would place us in
trouble with states with Stark energies around the maximum of the
atom in static field potential $E_{max}$, were it not for the sharp
increase there of the decay rates from practically zero to
practically infinity over a range of two or three states. This
makes me confident that the errors connected with the use of the
wrong states will concern only a few states close to the ionization
border which never in our simulations get significantly populated.
The continuum above the static field ionization threshold gives no
cause for worry: for most of my simulations the uppermost level is
$n_{max} =221$. This gives us about $60$ states above the static
field ionization threshold, extending in energy about $5.6*10^{-5}
\hspace{.1in}a.u.$ above that threshold. For the frequencies of my
simulations ($\omega\lesssim 2.7*10^{-6} a.u.$) this span in energy
is certainly sufficient; moreover, due to the big decay rates of
the states of this ``pseudo-continuum", the system will not be able
to resolve the individual states. More serious cause of worry is
that my procedure actually cuts away pieces of the ``double well"
states and this might cause interference effects altering the final
result of the simulation. On the other hand such interference
effects should be highly sensitive to changes of the basis and extensive
tests show this not to be the case.

When the microwave field is switched on, some indication on the
validity of the above approximation can be extracted from the
quasienergy curves. Ref. \cite{ref19} has shown that the
quasienergies calculated on a truncated basis are subject to the
unphysical constraint that their sum is a constant independent from
$F$ but dependent on the number of states in the basis. Whatever
our choice of basis, certain curves (at the top of the basis) are
going to be completely wrong. Now, with increasing microwave field
some of the quasienergy states move up and some down;  the wrong
states are therefore bound to undergo a series of avoided
crossings. Since the coupling among the states increases with the
microwave field strength, the wrong curves soon or later interact
(have a pronounced avoided crossing) with other curves. The avoided
crossings of these other curves for fields above the field at which
this interaction takes place therefore become in their turn
unreliable and so on, to the point that all the quasienergies of
interest bear no resemblance to those of the complete problem
\cite{ref44}. Often it is still possible to extract some
information from  the ``wrong" sections of the curves, but it will
be mostly qualitative. There are some rules of the thumb to
recognize the completely wrong curves. Levels close to the top of
the basis are always suspect. Moreover when the nearest neighbour
free atom (or, in our case, atom in static field) energy spacing is
more than $\omega\hbar$ one expects \cite{ref19} the quasienergy
curves to bend down and viceversa. Curves which do not follow this
pattern are usually suspect. The most reliable way of finding the
wrong curves is still, on the other hand, to perform test
calculations on different bases. For my calculations the free atom
basis used was the same I used for the pulsed simulations. For
frequencies of interest the lowest level $n_{min} = 30$ (and those
just above it) has no wide avoided crossings and is practically at
the same quasienergy over the entire range of field strengths
spanned by our simulations. Tests where I included all the states
down to n=1 show that all levels with lower quantum numbers behave
the same way.  Tests where I reduced the upmost level from
$n_{max}=221$ to $n_{max} =98$ show that, for the microwave
frequencies of interest here, the behaviour of the levels near
$n_0$ are very similar over the range of  field strengths spanned
by my simulations. On the other hand the avoided crossings of the
relevant states with the wrong curves at the top of the basis are
very narrow for the bigger basis but become much wider for the
smaller basis.

We have seen that the free atom basis somehow functions as a
Sturmian one does in absence of a static field, reproducing the
resonances above the classical static field ionization threshold;
but we have also seen that truncation of this basis means the
presence of an artificial second potential well with unphysical
finite number of discrete states in it. One should therefore wonder
about the impact of these levels on the levels of interest. Curves
corresponding to states deep in the second ``well" tend to have
very little interaction with any other state and higher second
``well" states tend to have significant avoided crossings only with
each other (avoided crossings wide enough that for our pulse
lengths they are not crossed diabatically and some population
exchange takes place \cite{ref33}). For the range of microwave
field strengths of interest, the only ``second well" states having
(significant) avoided crossings influencing the ``bound states" we
are interested in are those close to the classical static field
ionization threshold.

Another parameter which can create serious problems is the
integration step $\Delta t$: it must be small enough so that the
spurious frequencies $2\pi k/\Delta t$  induced by the time
discretization \cite{ref14} do not cause transitions between the
basis states. Approximating the basis upper limit with the free
atom continuum, this means that any $\Delta t < 4\pi n_{min}^2$ is
sufficient to inhibit spurious transitions even between the very
little populated extremes of the basis. For $n_{min}
=30$, the above condition is $\Delta t < 10^4 a.u.$. For most of my
simulations $\Delta t = 5\cdot10^3 a.u.$.

\section{}

\subsection{An evaluation of the birth and growth of the secondary island
chains using the classical pendulum approximation for the principal primary resonance island \cite{ref40}}
\label{app:duea}

When $F(t)$ is a constant, one procedure often used to study the
Hamiltonian (1) is to locally approximate it with an integrable one
around any resonant action $I_r$ defined by $\omega I_r^3 /(1-3F_s
I_r^4)= s$ where $s$ is an integer \cite{ref12}. The resulting
Hamiltonian is that of a pendulum describing the slow motion inside
the resonance island; for the principal primary resonance $s=1$, it
reads
\begin{equation}
H^{(2)}  =  - {\frac {\beta x^2} 2} + \alpha \cos{\varphi}
\label{eqb1}
\end{equation}
where
\begin{eqnarray}
\beta=3(1+F_SI_r^4), \hspace{.3in}\alpha = 0.325 F_0,\nonumber\\
\varphi = \theta - \omega t,\hspace{1.0in}\nonumber
\end{eqnarray}
and
\begin{eqnarray}
x= {\frac I {I_r}} -1\nonumber
\end{eqnarray}
The {\it action angle variables} of the Hamiltonian (\ref{eqb1})
itself are well known. For rotating motion outside the separatrix
the (rescaled) action and angle are \cite{ref38}
\begin{eqnarray}
J = {\frac 4 \pi} \left({\frac \alpha \beta}\right)^{1/2}
{\frac{{\bf E}(R)} {R}},\nonumber\\
\varphi_J = \pm \pi {\frac {F(\varphi/2, R )} {{\bf K}(R)}}
\label{eqb4}
\end{eqnarray}
where
\begin{eqnarray}
R = \left({\frac 2 {1 - H/ \alpha}}\right)^{1/2} = {\frac 1
{|sin(\Phi/2)|}},\nonumber
\end{eqnarray}
$\Phi$ being the (half) amplitude in $\varphi$ of the oscillations.
${\bf E}$ and ${\bf K}$ are complete elliptic integrals
\cite{ref38}, $F$ is the incomplete elliptic integral corresponding
to ${\bf K}$, and the $+$ sign refers to $x>0$. Analogously - for
librating motion inside the separatrix - \cite{ref38}
\begin{eqnarray}
J = {\frac 8 \pi}
\left({\frac \alpha \beta}\right)^{1/2}\left[{{{\bf E}(1/R)
- \frac{(R^2-1){\bf K}(1/R)} {R^2}}}\right],\nonumber\\
\varphi-J= \pm \pi {\frac{R F(\varphi/2,R)} {2 {\bf K}(1/R)}}\hspace{.7in}
\label{eqb3}
\end{eqnarray}
The frequency of motion in this latter case, known as {\it ``slow
motion frequency"} \cite{ref39}, is:
\begin{equation}
\Omega \equiv  |d\varphi_J /dt| = |\partial H^{(2)}/ \partial J| =  \pi \left({\alpha \beta}\right)^{1/2} /2 {\bf K}(1/R)
\label{eqb8a}
\end{equation}
which is maximum for $1/R = 0$ where it becomes the usual ``small
oscillations" frequency
\begin{equation}
\Omega_{max} =  (\alpha\beta)^{1/2}.
\label{eqb8b}
\end{equation}
Since there is for every $F$ a maximum slow frequency
$\Omega_{max}$, the resonance condition for a secondary resonance
with rotation number $p/q$ can be verified only if
\begin{equation}
\Omega_{max} > p\omega/q = p(1 -3F_S I_r^4 )/qI_r^3.
\label{eqb9a}
\end{equation}
The above condition for the existence of a $p/q$ secondary
resonance zone can be written in the form of a condition on the
microwave field strength:
\begin{equation}
FI_r^4 > F_{p/q}I_r^4 \equiv {\frac{(1 -3F_S I_r^4 )^2 p^2} {0.975
q^2(1 + F_SI_r^4)}} \cong {\frac{(1-7F_S I_r^4 )p^2} {0.975 q^2}}
\label{eqb9b}
\end{equation}
For $F < F_{p/q}$ the secondary resonance zone does not exist and
it appears for $F = F_{p/q}$ at  the center of the primary
resonance zone, $\Omega_{max}$ being the pendulum libration
frequency there. Contrary to the primary resonance zones, it is
clear that since $\Omega = \Omega (J,F)$, the secondary resonance zones do move
in phase space. To see how they move we can combine (for $F >
F_{p/q}$) the $p/q$ resonance condition (\ref{eqb9a}) with eq. (\ref{eqb8a})
expanded in powers of $1/R=|sin(\Phi/2)|$; this gives us the
distance in $\varphi$ of the secondary island chain from the
primary fixed point:
\begin{equation}
\Phi \cong (8 \Delta F/ F_{p/q})^{1/2}
\label{eqb10}
\end{equation}
where $\Delta F = F - F_{p/q}$. We thus see that for every $p/q$
the resonance starts for $F = F_{p/q}$  - called the $p/q$
bifurcation field - at the fixed point itself ($\Phi =0$) and then
moves outwards toward higher $\Phi$'s with increasing $F$.

Eqs. (\ref{eqb9b}) and (\ref{eqb10}) have been obtained in pendulum
approximation, we therefore expect them to work only up to moderate
fields; their failure for low $q$ integer resonances - which appear
only at high microwave fields - is particularly evident for eq.
(\ref{eqb10}) where the exponent $1/2$ does not fit the numerical
results for $q\leq 4$ \cite{ref9}. Another evaluation of the
bifurcation fields $F_{p/q}$ can be obtained \cite{ref40} from a
stability analysis of the Kepler map: an approximate map for the
classical dynamics derived in Ref. \cite{ref12}. In the map the
fixed point at the center of the principal primary resonance
becomes unstable after the $p/q=1/2$ bifurcation point. This is
confirmed by my numerical simulations for the complete system which
show no trace of the primary stable fixed point above the $F_{1/2}$
bifurcation point. The fractional deviation of the Kepler map
bifurcation points from the numerical ones is approximately
constant: they are always $25\%$ below the numerical ones, while
the pendulum approximation bifurcation points are always above the
numerical ones and their deviation increases from about $5\%$ for
$q=7$ and $6$ to about $128\%$ for $q=2$. The introduction of a
static field $F_S
= 8 V/cm$ reduces in our case the bifurcation points by about $20\%$.
Again, in all cases the results in pendulum approximation are
higher than the numerical ones and the results from the Kepler map
lower; but while the deviation of the pendulum approximation
results increases from about $15\%$  at $q=6$ to about $150\%$ at
$q=2$, the deviation of Kepler map results from the numerical ones
remains about $15\%$ from $q=6$ to $q=2$. Since in absence of a
static field the deviation was about $25\%$; this suggest that the
linear approximation of the static field I used is an
over-evaluation.

\subsection{quantum pendulum approximation and WKB quantization \cite{ref40}}
\label{app:dueb}

Following Ref. \cite{ref20} and \cite{ref29}, the quasienergies of
the Hamiltonian (1) can be approximated by
\begin{equation}
\varepsilon^{(1)}_k = (E_R + W_k ) (mod \hspace{.1in}\omega\hbar).
\label{eqb14}
\end{equation}
where $E_R$ is the eigenvalue of the (classically) resonant state
$|R>$ of the Hamiltonian for the atom in the static field alone;
and $W_k$ denote the eigenvalues of the matrix $H^{(1)}_{n,m}$
defined as follows:
\begin{eqnarray}
H^{(1)}_{n,n} = [E_n - E_R - \omega(n-R)],\nonumber\\
H^{(1)}_{n,n+1} = F \left<{R|z|R+1}\right> /2
\label{eqb13b}
\end{eqnarray}
While the diagonalization of the Floquet operator gives us no
natural ordering of the quasienergies, the energies $W_k$ can be
ordered. A descending order in energy with $k$ starting from zero
allows us to identify the index $k$ as the quantized version of the
pendulum action $J$. Moreover, since eq. (\ref{eqb14}) gives
energies folded into the first Brillouin zone, the level crossings
appearing as a consequence of this folding are actual crossings.
Eq. (\ref{eqb14}) has been shown to give for quantum numbers close
to $R$ quasienergy curves remarkably similar to those of the full
problem \cite{ref29} for $n_0=60$ and rescaled fields up to about
$F_0=0.04$. I have numerically calculated the ``quasienergies"
(\ref{eqb14}) for the parameters of Fig. \ref{fig1}; the matrix
elements $H^{(1)}_{n,m}$ between the Stark states were calculated
by projecting the Stark states on the free atom basis $n\in
[33,94]$.The result is shown as the dotted curves in Fig.
\ref{figb1}. For comparison, I also plotted - as big dots
connected by dark lines - the result of a WKB quantization of
easily calculated approximations of the constant action $J_0$
curves of energy versus microwave field strength $F_0$ implicitly
defined by eqs. (\ref{eqb3}) and (\ref{eqb4}). The approximations
made were the following: The zero action curve is the energy of the
fixed point: $H = \alpha$. Near $J_0 =0$ we can expand ${\bf
K}(1/R)$ and ${\bf E}(1/R)$ in powers of $1/R$ (to the fourth order
\cite{ref38}) and then solve for $H$ the resulting approximation of
eq. (\ref{eqb3}). For a given action $J_0$ this approximation will
get better and better for increasing $\alpha$: in the high field
limit equidistance in action will therefore mean equidistance in
energy and the system will behave essentially like a harmonic
oscillator with frequency $\Omega= (\alpha\beta)^{1/2}$. A similar
procedure is used in the opposite limit ($\alpha
\rightarrow 0$) starting from eq. (\ref{eqb4}). At the separatrix ($H = -
\alpha$), there is a discontinuity in $J$: $J = 4(\alpha/\beta)^{1/2}/\pi$
outside and $J= 8(\alpha/\beta)^{1/2}/\pi$ inside; but, from both
sides, approaching the separatrix, $dH/d\alpha$ tends to $-1$ so
that the $H(\alpha)$ curves join smoothly ($C^1$ at least). Since
their slope at that point is that of the separatrix itself, we also
know we have an inflection point there. Finally the curve
$H(\alpha)$ has a minimum given by the condition $2{\bf E}(1/R)
= {\bf K}(1/R)$ which solved with the help of the numerical tables \cite{ref37}
gives us ${\bf K}(1/R) = 2.3211$ and $J = 8
(\alpha/\beta)^{1/2}(-H/a) {\bf E}(1/R)/\pi \cong 1.1128
(3\alpha/\beta)^{1/2}$. With the above approximations and using the
WKB $n$-quantization outside the classical resonance zone and the
$k$-quantization inside, we have:

Low field behaviour:
\begin{eqnarray}
{\frac{\varepsilon_k} \omega} \cong \left[{\frac{- |H_0|
-{\frac{\alpha^2} {8| H_0|}}}  \omega}\right] mod(1)\hspace{.4in}\nonumber\\
= \left[{{\frac{\varepsilon_k(0)} \omega}- {\frac{\left({\frac{0.325 F_0 I_r^2} {n_0\omega_0}}\right)^2 } {\frac{8|\Delta_k(0)|} \omega}}}\right] mod(1),\nonumber
\end{eqnarray}
Minimum:
\begin{eqnarray}
F_0= {\frac{1+ F_s^0(I_r/n_0)^4} {0.325}} \left[{\frac{(k
+1/2)n_0^2} {1.1128 I_r^3}}\right]^2\nonumber\\ {{\varepsilon_k}
\over\omega} \cong  \left[{\frac{\varepsilon_r} \omega} -  0.6522
{\frac \alpha \omega}\right]mod(1)\hspace{.5in}\nonumber\\
= \left[{\frac{\varepsilon_r} \omega} -  0.6522{\frac{0.325F_0I_r^2} {n_0\omega_0}}\right]mod(1)
\label{eqb15}
\end{eqnarray}
High field behaviour:
\begin{eqnarray}
{\frac{\varepsilon_k} \omega} \cong \left\{{\frac{\varepsilon_r}
 \omega} + {\frac\alpha \omega}\left[{9 - 8\left(1 +{\frac{J_0
(3/\alpha)^{1/2}} 4}\right)^{1/2}}\right]\right\}mod(1)\nonumber\\
=\left\{{\frac{\varepsilon_r} \omega} + {\frac{0.325F_0I_r^2} {n_0\omega_0}}\left[{9 - 8\left(1+{\frac {\mathcal{R}} 4}\right)}\right]^{1/2}\right\}mod(1)\nonumber
\end{eqnarray}
where
\begin{eqnarray}
{\mathcal R}={\frac{(k +1/2)n_0^2} {I_r^3}}\left({\frac{3[1+ F^s_0(
I_r/n_0)^4]} {0.325F_0}}\right)^{1/2}.\nonumber
\end{eqnarray}
In Fig. \ref{figb1} the resonant action is $I_r = 69$ and the zero
microwave field parameters $\varepsilon_k(0)$ and $|\Delta_k(0)|= -
\Delta_k(0) = W_k(0)= \varepsilon_r - \varepsilon_k(0)$ used in the
construction are given in table \ref{table2} (in atomic units). The
agreement is quite good both at high and low fields; the shift of
the high field curves increases with $k$ but for each of these
curves it appears to slowly decrease with increasing field. The
inflection points, which classically are right at the separatrix,
are at the center of the transition region between the two
quantizations: to their left the states are essentially modulated
Stark states while to their right they take the character of
resonant states. In our semiclassical treatment we can therefore
only say that they fall somewhere between the $n$-quantized point
and the $k$-quantized one (circled points connected by segments in
Fig. \ref{figb1})

\subsubsection{Semiclassical scaling of the quasienergy curves \cite{ref40}}

Reference \cite{ref41} suggests an approximate semiclassical
scaling law:
\begin{equation}
\varepsilon_n(F) / \omega^{2/3} \approx f_{n\omega^{1/3}} ( F/\omega^{4/3})
\label{eqb16}
\end{equation}
where the function $f$ depends only on classically rescaled
quantities. Eq. (\ref{eqb16}) connects  quasienergy curves at
different frequencies: to get the same function $f_{n\omega^{1/3}}$
for a different quantum number $n$, $\omega$ has to be changed. The
agreement of my classical pendulum approximation evaluations with
the quantum curves confirms for the cases studied this hypothesis,
but only up to a certain point: while my evaluation for the low
field behaviour, the top equation in (\ref{eqb15}), scales exactly
according to eq. (\ref{eqb16}), the evaluations for the minimum and
the high field behaviour depend not only on $n$ but also on $(k +
1/2)/n$, as can be seen from the two bottom eqs. in (\ref{eqb15}).


\newpage

\begin{table}
\begin{eqnarray} \hline\hline\nonumber
\nonumber\\
  Figure & \hspace{0.1in} K & \hspace{0.2in} Population \hspace{0.1in}(\% \hspace{0.1in}of \hspace{0.1in}total \hspace{0.1in}probability)\nonumber\\
\hline\nonumber\\
  10.a & 0 & \hspace{0.2in}57 \nonumber\\
    & 1 & \hspace{0.2in}16\hspace{0.05in} \overrightarrow{(7)}\hspace{0.05in} 15 \hspace{0.05in}\overrightarrow{(7*)} \hspace{0.05in} 13\nonumber\\
    & 2 & \hspace{0.2in}15 \hspace{0.05in} \overrightarrow{(8)}\hspace{0.05in} 14 \hspace{0.05in}\overrightarrow{(8*)} \hspace{0.05in} 12\nonumber\\
    & 3 & \hspace{0.2in}10 \nonumber\\
  10.b & 0 & \hspace{0.2in}56\hspace{0.05in} \overrightarrow{(7)}\hspace{0.05in} 55\hspace{0.05in} \overrightarrow{(6*)} \hspace{0.05in} 54 \nonumber\\
    & 1 & \hspace{0.2in}15\hspace{0.05in} \overrightarrow{(7)}\hspace{0.05in} 14\hspace{0.05in} \overrightarrow{(6)}\hspace{0.05in} 11\hspace{0.05in} \overrightarrow{(7*)} \hspace{0.05in}10\nonumber\\
    & 2 & \hspace{0.2in}15\hspace{0.05in} \overrightarrow{(8)}\hspace{0.05in} 14\hspace{0.05in} \overrightarrow{(7)}\hspace{0.05in} 8\hspace{0.05in} \overrightarrow{(8*)}\hspace{0.05in} 6\nonumber\\
    & 3 & \hspace{0.2in}12\hspace{0.05in} \overrightarrow{(7)}\hspace{0.05in} 2\nonumber\\
  10.c & 0 & \hspace{0.2in}53\hspace{0.05in} \overrightarrow{(5)}\hspace{0.05in} 49\hspace{0.05in} \overrightarrow{(5*)} \hspace{0.05in} 47  \nonumber\\
    & 1 & \hspace{0.2in}13\hspace{0.05in} \overrightarrow{(6)}\hspace{0.05in} 12\hspace{0.05in} \overrightarrow{(5)}\hspace{0.05in} 9\hspace{0.05in} \overrightarrow{(5*)}\hspace{0.05in} 7\nonumber\\
    & 2 & \hspace{0.2in}15\hspace{0.05in} \overrightarrow{(8)}\hspace{0.05in} 14\hspace{0.05in} \overrightarrow{(6)} \hspace{0.05in}11\hspace{0.05in} \overrightarrow{(5)} \hspace{0.05in}3\hspace{0.05in} \overrightarrow{(5*)} \hspace{0.05in} 1\nonumber\\
    & 3 & \hspace{0.2in}14\hspace{0.05in} \overrightarrow{(7)}\hspace{0.05in} 12\hspace{0.05in} \overrightarrow{(6)} \hspace{0.05in}9\hspace{0.05in} \overrightarrow{(6*)}\hspace{0.05in} 5  \nonumber\\
  10.d & 0 & \hspace{0.2in}50\hspace{0.05in} \overrightarrow{(5)}\hspace{0.05in} 47\hspace{0.05in} \overrightarrow{(4)} \hspace{0.05in} 41\hspace{0.05in} \overrightarrow{(3)} \hspace{0.05in}23\hspace{0.05in} \overrightarrow{(X*)}\hspace{0.05in} 20\nonumber\\
    & 1 & \hspace{0.2in}12\hspace{0.05in} \overrightarrow{(4)}\hspace{0.05in} 3\nonumber\\
    & 2 & \hspace{0.2in}14\hspace{0.05in} \overrightarrow{(5)}\hspace{0.05in} 8\hspace{0.05in} \overrightarrow{(4)}\hspace{0.05in} 1\nonumber\\
    & 3 & \hspace{0.2in}15\hspace{0.05in} \overrightarrow{(6)}\hspace{0.05in} 12\hspace{0.05in} \overrightarrow{(5)}\hspace{0.05in} 5\nonumber\\ \hline\hline\nonumber
\end{eqnarray}
\caption{The change of population on the four central resonance states in
Figures \ref{fig10a}-\ref{fig10d} at the relevant avoided
crossings. The populations are in percent of the total unit
population. The $q$ for each crossing is given in parenthesis, an
asterisc follows crossings on the falling edge of the pulse. The
$x$ denotes a $p=2$, $q=9$ crossing.}
\label{table1}
\end{table}

\begin{table}
\begin{eqnarray} \hline\hline\nonumber
\nonumber\\
  k  \hspace{0.2in} n  \hspace{0.2in} E_k(0)/\omega  \hspace{0.2in} |\Delta_k(0)|/\omega \nonumber\\
\hline\nonumber\\
  0 \hspace{0.2in} 69 \hspace{0.3in} 0.24370 \hspace{0.3in} 0 \hspace{0.3in}\nonumber\\
  1 \hspace{0.2in} 70 \hspace{0.3in} 0.23218 \hspace{0.3in} 0.0115 \nonumber\\
  2 \hspace{0.2in} 68 \hspace{0.3in} 0.20187 \hspace{0.3in} 0.0418 \nonumber\\
  3 \hspace{0.2in} 71 \hspace{0.3in} 0.17168 \hspace{0.3in} 0.0720 \nonumber\\
  4 \hspace{0.2in} 67 \hspace{0.3in} 0.10540 \hspace{0.3in} 0.1383 \nonumber\\
  5 \hspace{0.2in} 72 \hspace{0.3in} 0.06417 \hspace{0.3in} 0.1795 \nonumber\\
  6 \hspace{0.2in} 66 \hspace{0.3in} 0.95055 \hspace{0.3in} 0.2927 \nonumber\\
  7 \hspace{0.2in} 73 \hspace{0.3in} 0.91201 \hspace{0.3in} 0.3310 \nonumber\\
  8 \hspace{0.2in} 65 \hspace{0.3in} 0.73391 \hspace{0.3in} 0.5098 \nonumber\\
  9 \hspace{0.2in} 74 \hspace{0.3in} 0.71738 \hspace{0.3in} 0.5236 \nonumber\\
\hline\hline\nonumber
\end{eqnarray}
\caption{Quantum parameters for the semiclassical curves in Fig. \ref{figb1}. (From Ref. \cite{ref40})}
\label{table2}
\end{table}

\newpage
.

\newpage
.

\newpage
.

\begin{figure}[htbp]
\centering\epsfig{file=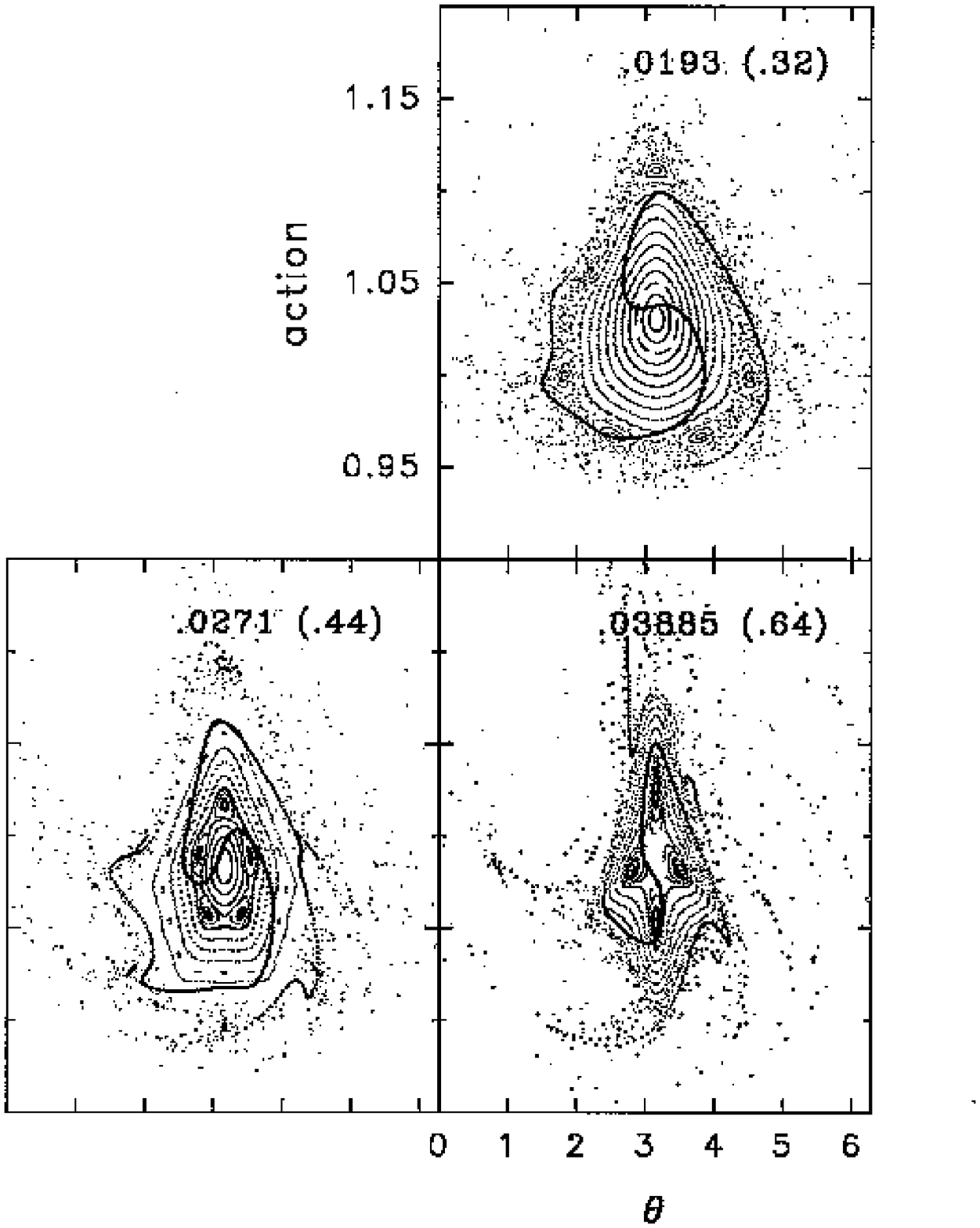,width=0.8\linewidth}
\end{figure}

\newpage
.

\begin{figure}[htbp]
\centering\epsfig{file=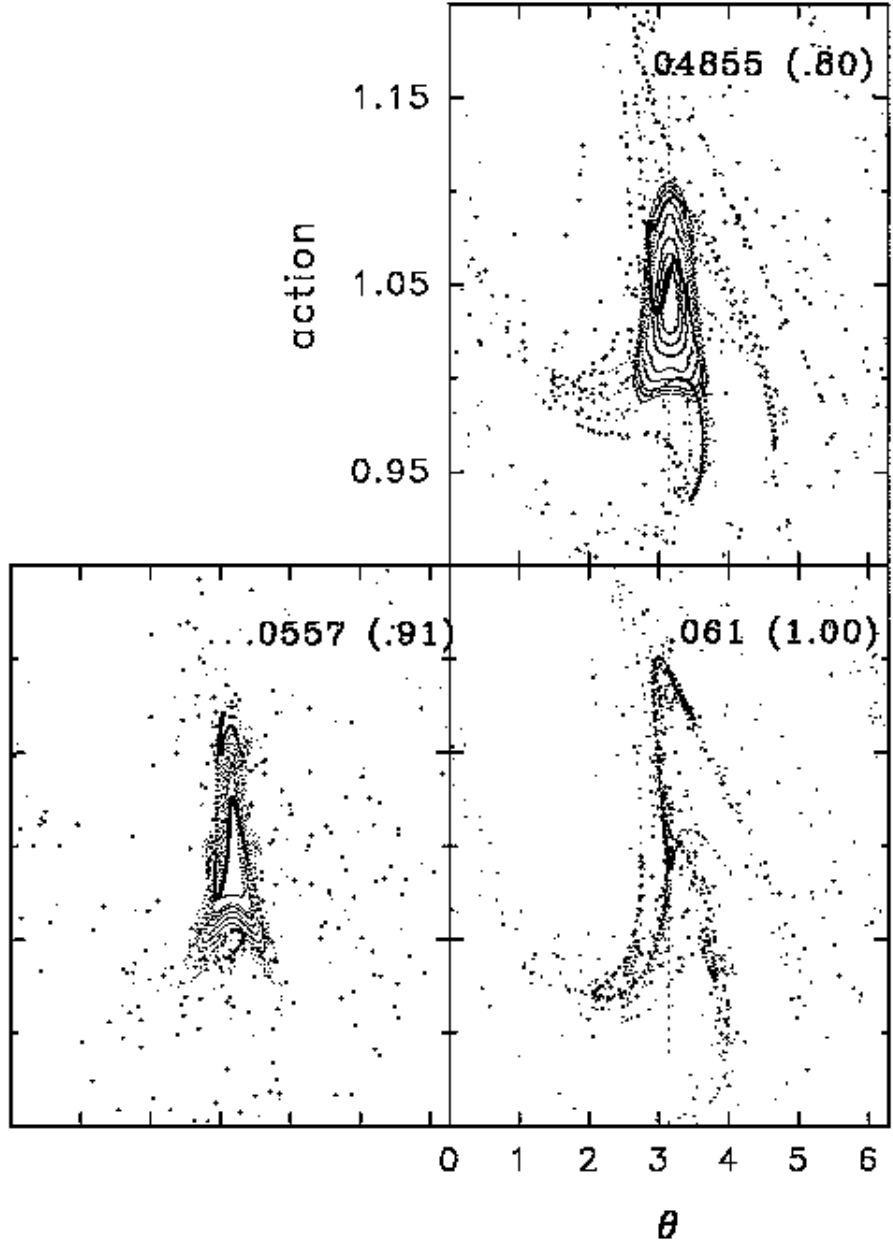,width=0.8\linewidth}
\caption{Comparison of snapshots from Fig. \ref{fig3d} ($F_0^{max} = 0.061$) with the surfaces of section
calculated at the values of the instantaneous $F_0(t)$; $F_0$ is
indicated at the top of each graph, its ratio to $F_0^{max}$ is
given in parenthesis. In the first and second snapshots the $q=7$
and $q=6$ chains respectively are the cause of the deformations of
the tails of the primary island whorl; in the third snapshot the
$q=4$ chain begins to act but the tendrils of the $q=5$ one are
still visible; so in the fourth the tendrils of the $q=4$ chain;
the fifth and sixth snapshots see the creation and development of
the $q=3$ tendrils (no stability islands are present for $q=3$).
(From Ref. \cite{ref40})}
\label{fig4d}
\end{figure}

\begin{figure}[htbp]
\centering\epsfig{file=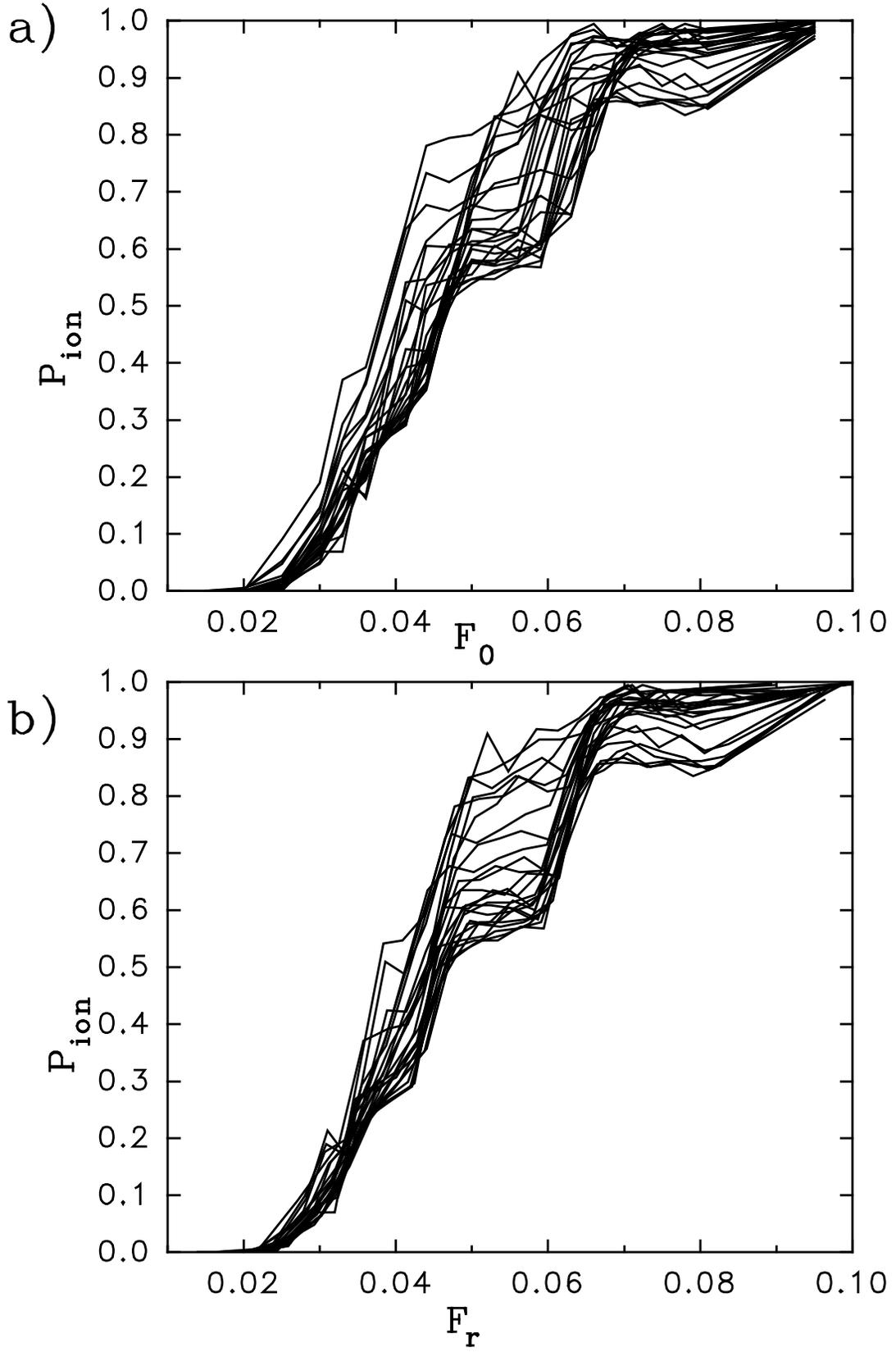,width=0.8\linewidth}
\caption{a) The classical probability for microwave
ionization of $n_0=69$ atoms plotted as a function of the rescaled
peak microwave field strength $F_0^{max}$. There is no alignment of
the ionization steps of the $25$ curves shown (at values of
$\omega'_0$ equispatiated between $\omega'_0=0.9325$ and
$\omega'_0= 1.0666$). b) The same but plotted as a function of peak
microwave field strength rescaled at the resonance $F_r^{max}$.
With this choice of the horizontal scale the ionization steps are
aligned. (From Ref. \cite{ref40})}
\label{fig5}
\end{figure}

\begin{figure}[htbp]
\centering\epsfig{file=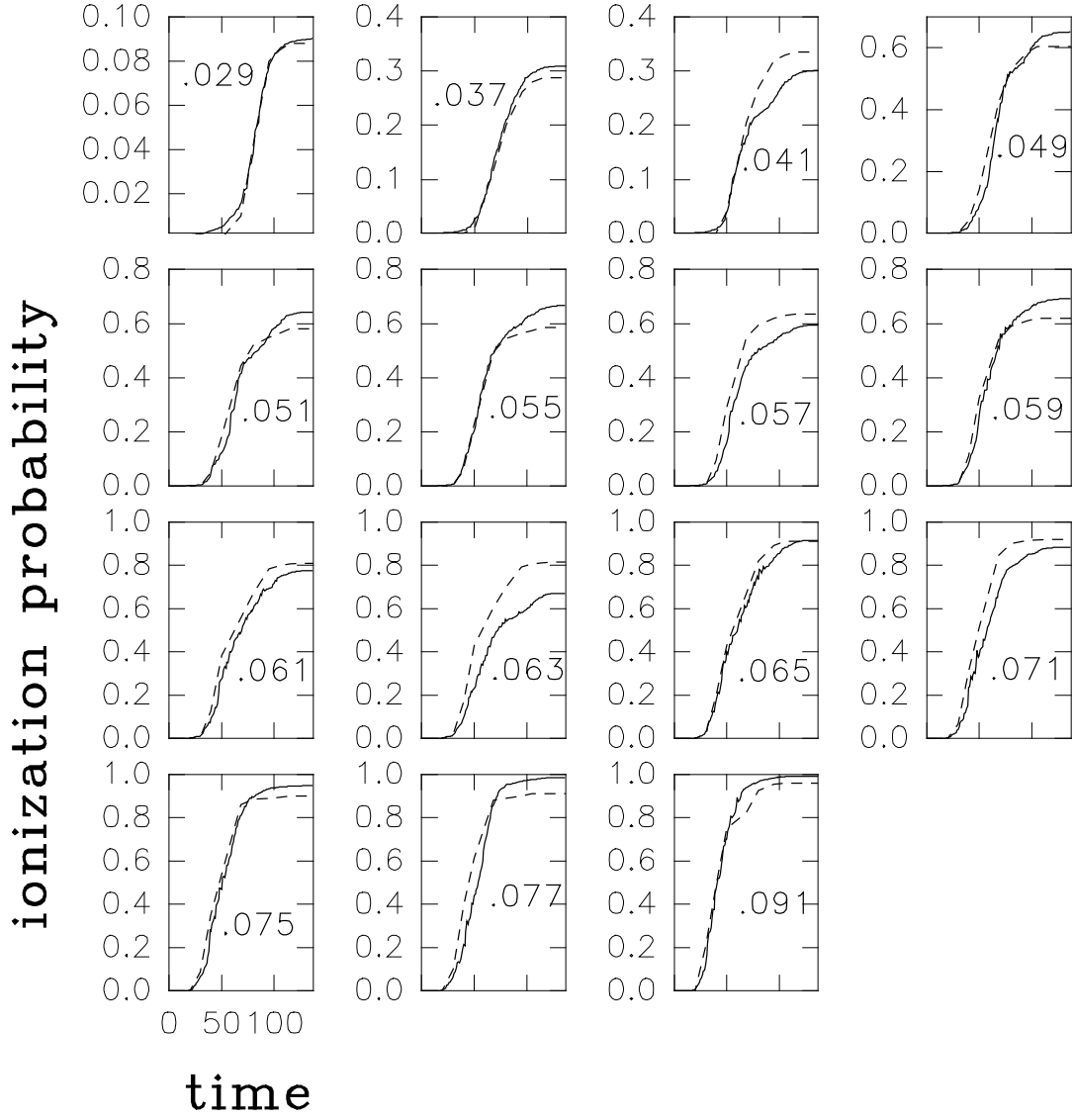,width=0.8\linewidth}
\caption{Comparison of quantum (full line) and classical (dashed line) time
evolutions for various values of $F_0^{max}$ (indicated next to
each plot). $F_0^S= 0.035266$ and $\omega'_0=0.9771$. (From Ref. \cite{ref40})}
\label{fig7}
\end{figure}

\begin{figure}[htbp]
\centering\epsfig{file=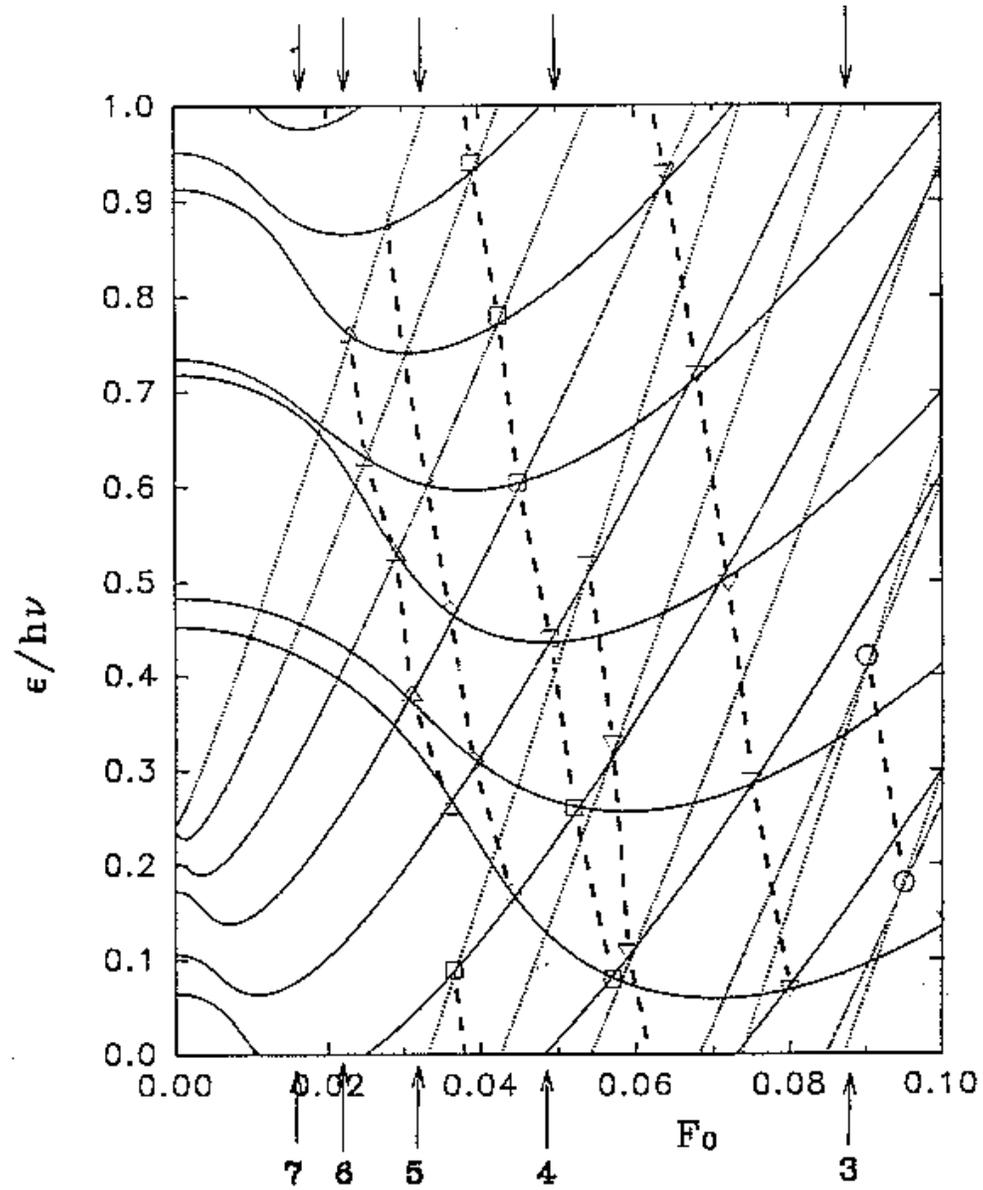,width=0.8\linewidth}
\caption{Pendulum approximation quasienergy curves for the case of Fig. \ref{fig1}.
The groups of crossings related to the most visible classical
secondary island chains are indicated by the same symbols as in
Fig. \ref{fig1}. Here the arrows mark the classical approximate bifurcation
fields eq (\ref{eqb9b}), labeled by $q$. (From Ref. \cite{ref40})}
\label{fig8}
\end{figure}

\begin{figure}[htbp]
\centering\epsfig{file=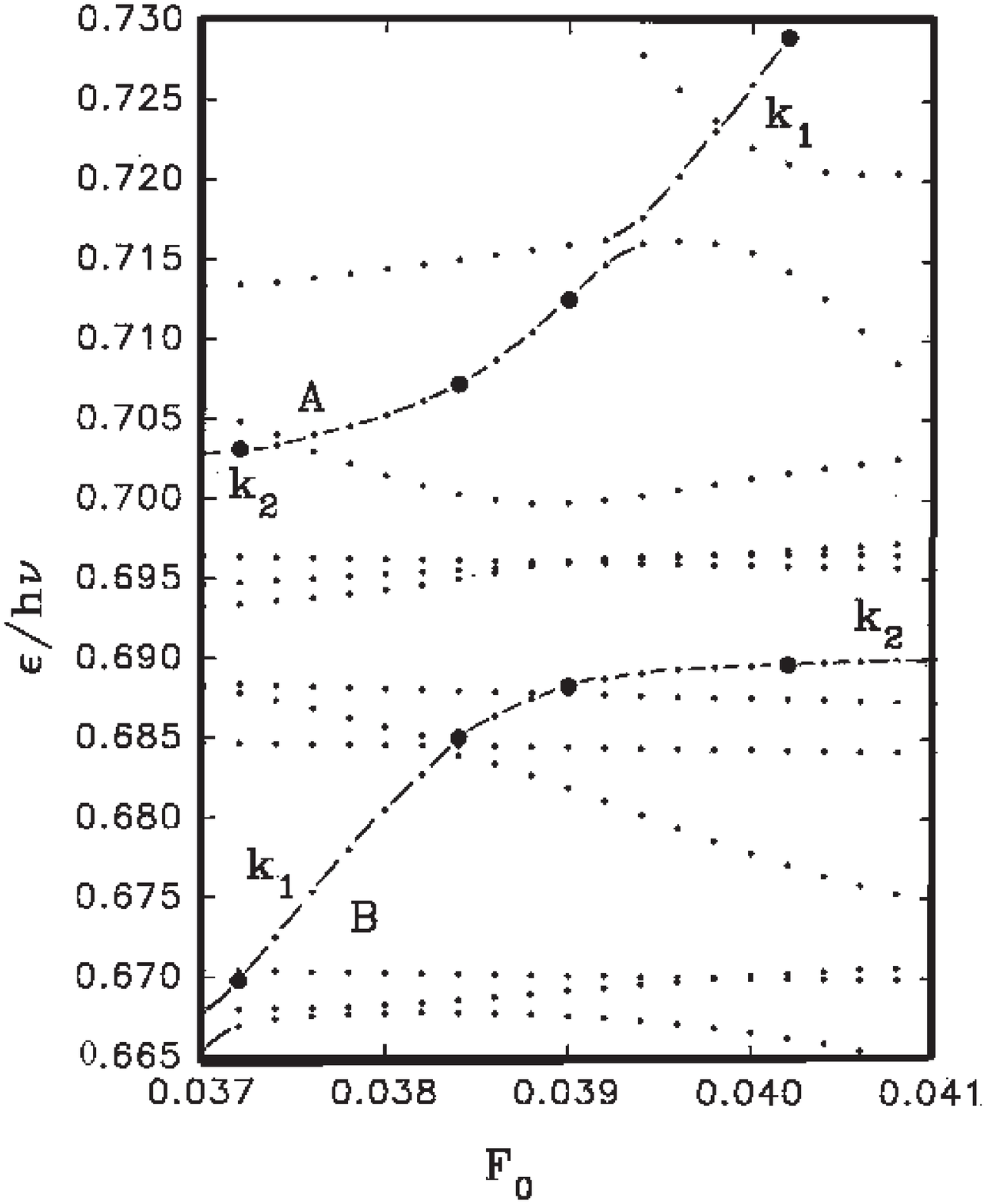,width=0.8\linewidth}
\end{figure}

\begin{figure}[htbp]
\centering\epsfig{file=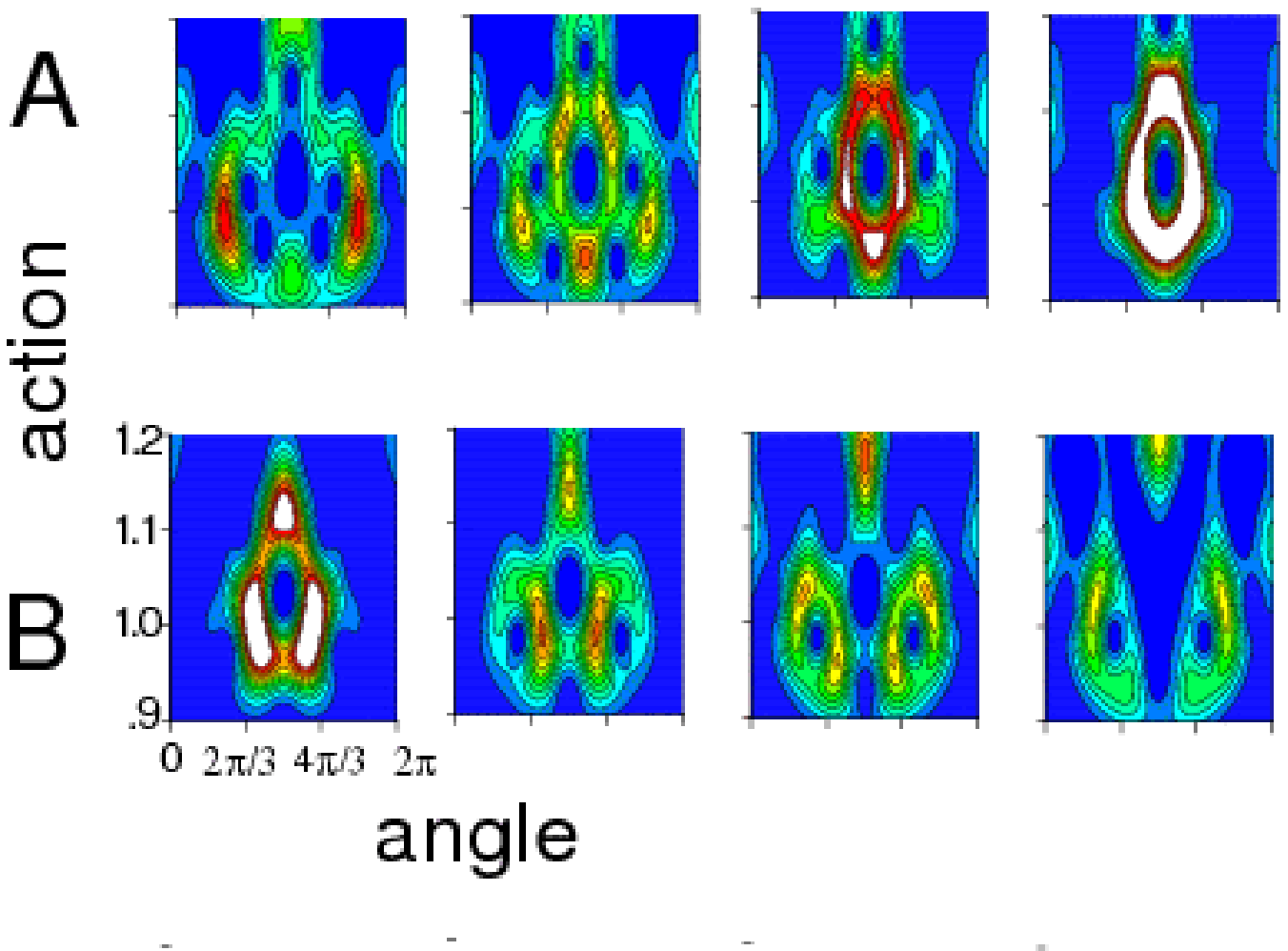,width=0.8\linewidth}
\caption{Detail of Fig. \ref{fig1} showing one of the avoided crossings marked in
Fig. \ref{fig1} and the Husimi functions of the (eigen-)states
undergoing this same avoided crossing: it is a wide crossing where
no other interacting level intervenes: the third $q=5$ crossing
($k_1=2$, $k_2=7$). The (adiabatic) states are labeled by letters
and numbers indicate the (diabatic) resonance quantization $k$.
Bigger dots mark on the quasienergy curves the points corresponding
to the Husimi functions displayed. Lighter shades of gray between
the level curves mean higher values of the Husimi functions. The
levels of the curves is the same for all plots; the highest peaks
in some of them are, as a result, out of range and appear as white
areas.}
\label{fig9a}
\end{figure}

\begin{figure}[htbp]
\centering\epsfig{file=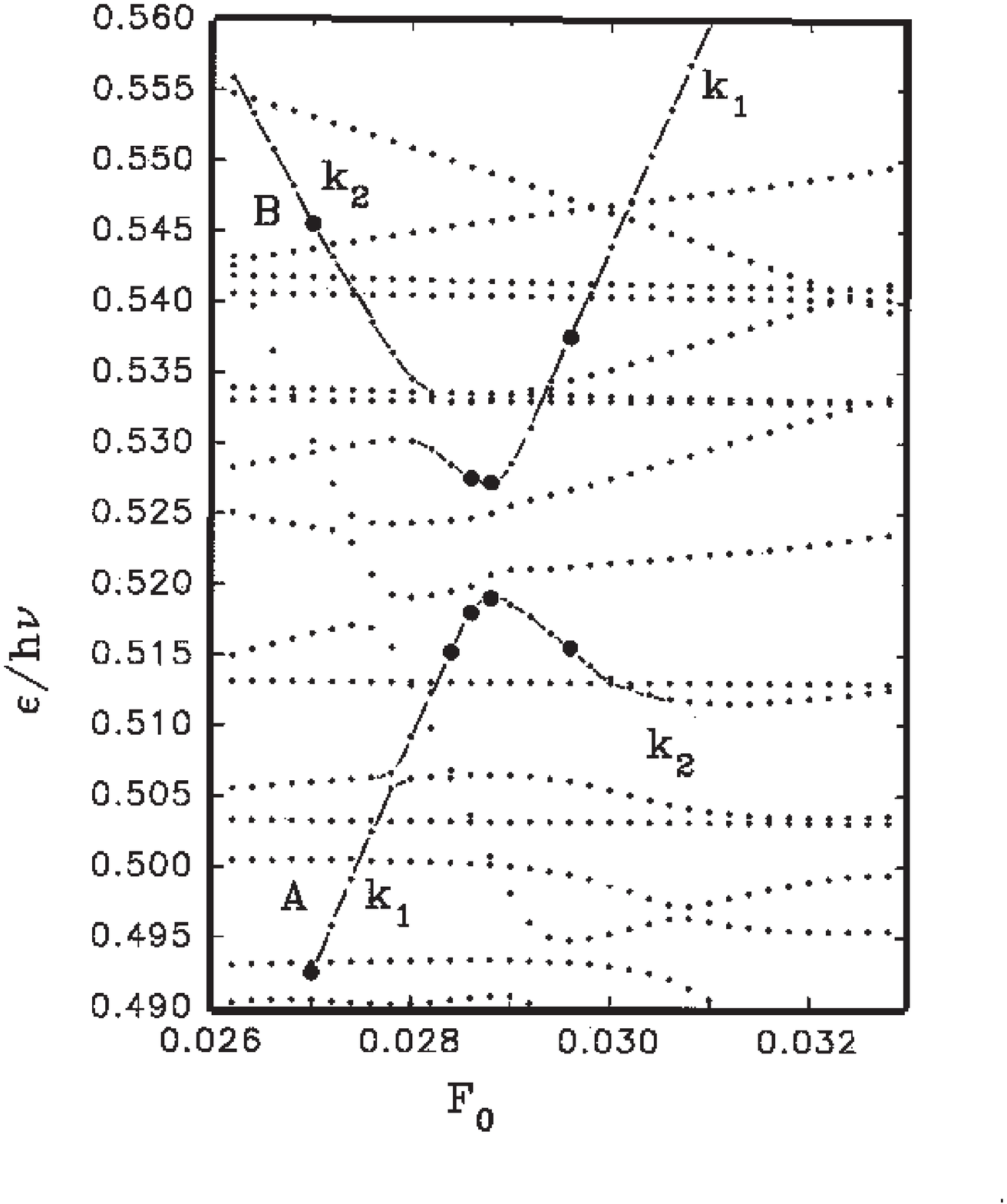,width=0.8\linewidth}
\end{figure}

\begin{figure}[htbp]
\centering\epsfig{file=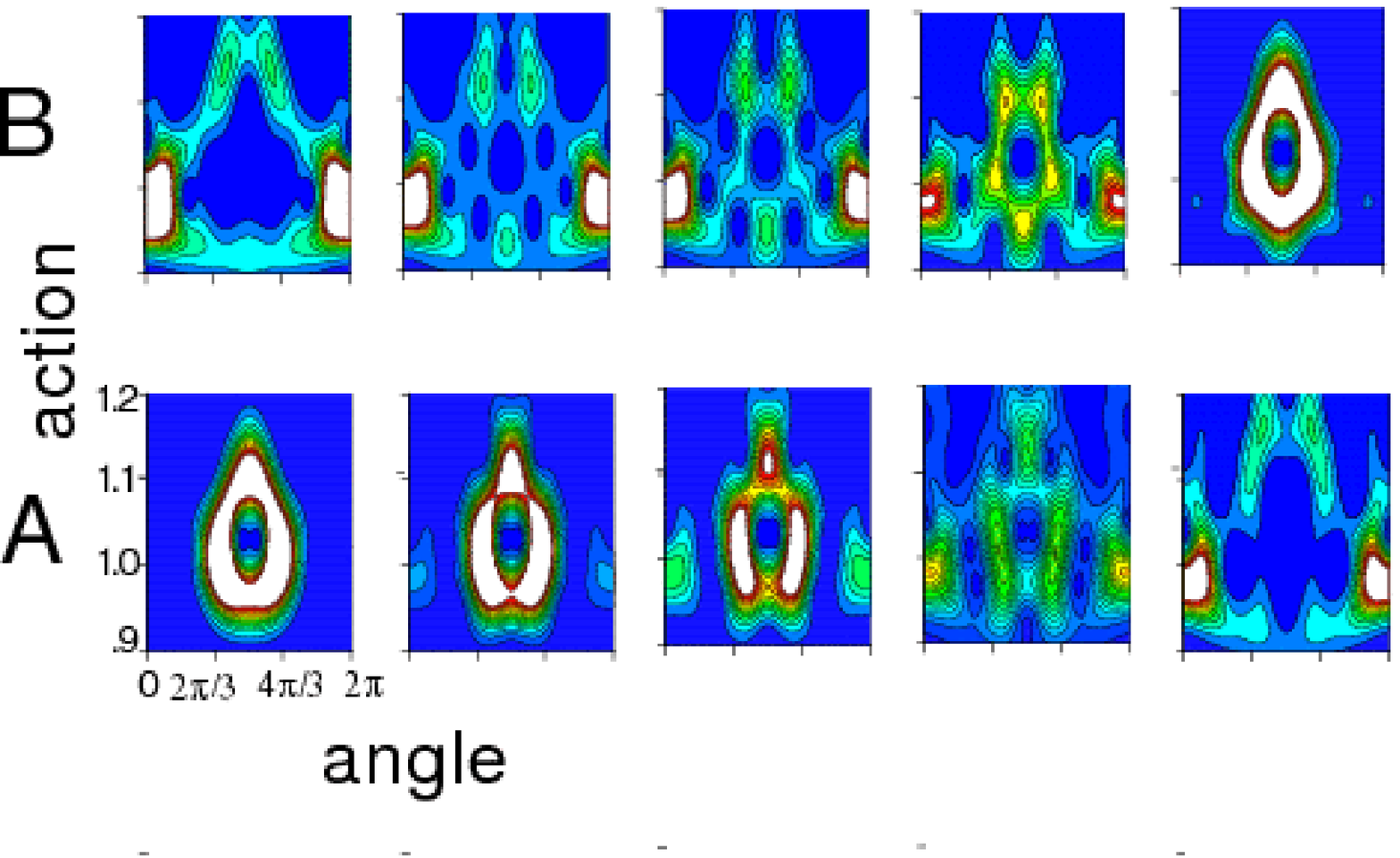,width=0.8\linewidth}
\caption{Detail of Fig. \ref{fig1} showing one of the avoided crossings marked in
Fig. \ref{fig1} and the Husimi functions of the (eigen-)states
undergoing this same avoided crossing: it is a narrow crossing
where no other level intervenes: the third $q=7$ crossing ($k_1=2$,
$k_2=9$). The (adiabatic) states are labeled by letters and numbers
indicate the (diabatic) resonance quantization $k$. Bigger dots
mark on the quasienergy curves the points corresponding to the
Husimi functions displayed. Lighter shades of gray between the
level curves mean higher values of the Husimi functions. The levels
of the curves is the same for all plots; the highest peaks in some
of them are, as a result, out of range and appear as white areas.}
\label{fig9b}
\end{figure}

\begin{figure}[htbp]
\centering\epsfig{file=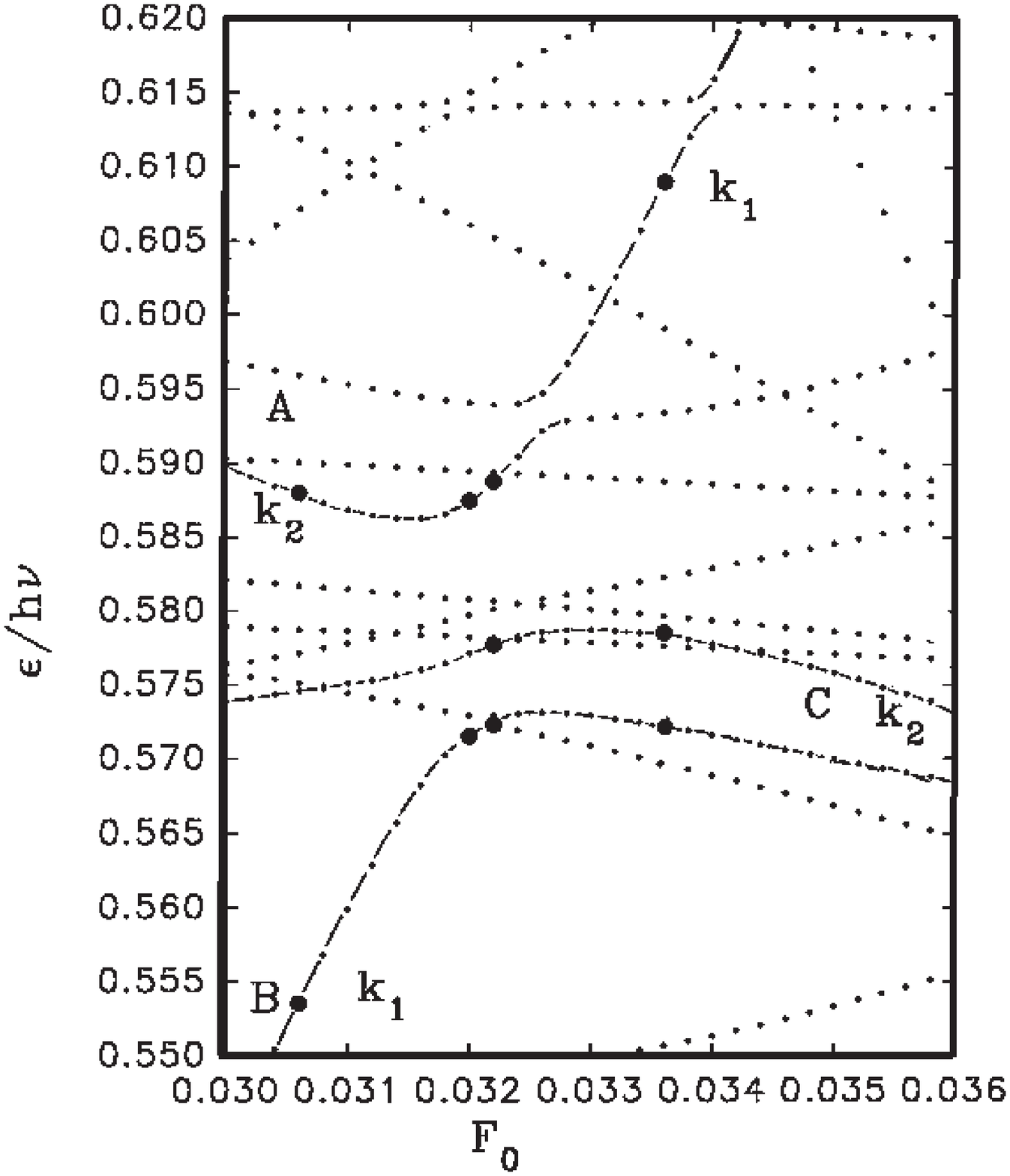,width=0.8\linewidth}
\end{figure}

\begin{figure}[htbp]
\centering\epsfig{file=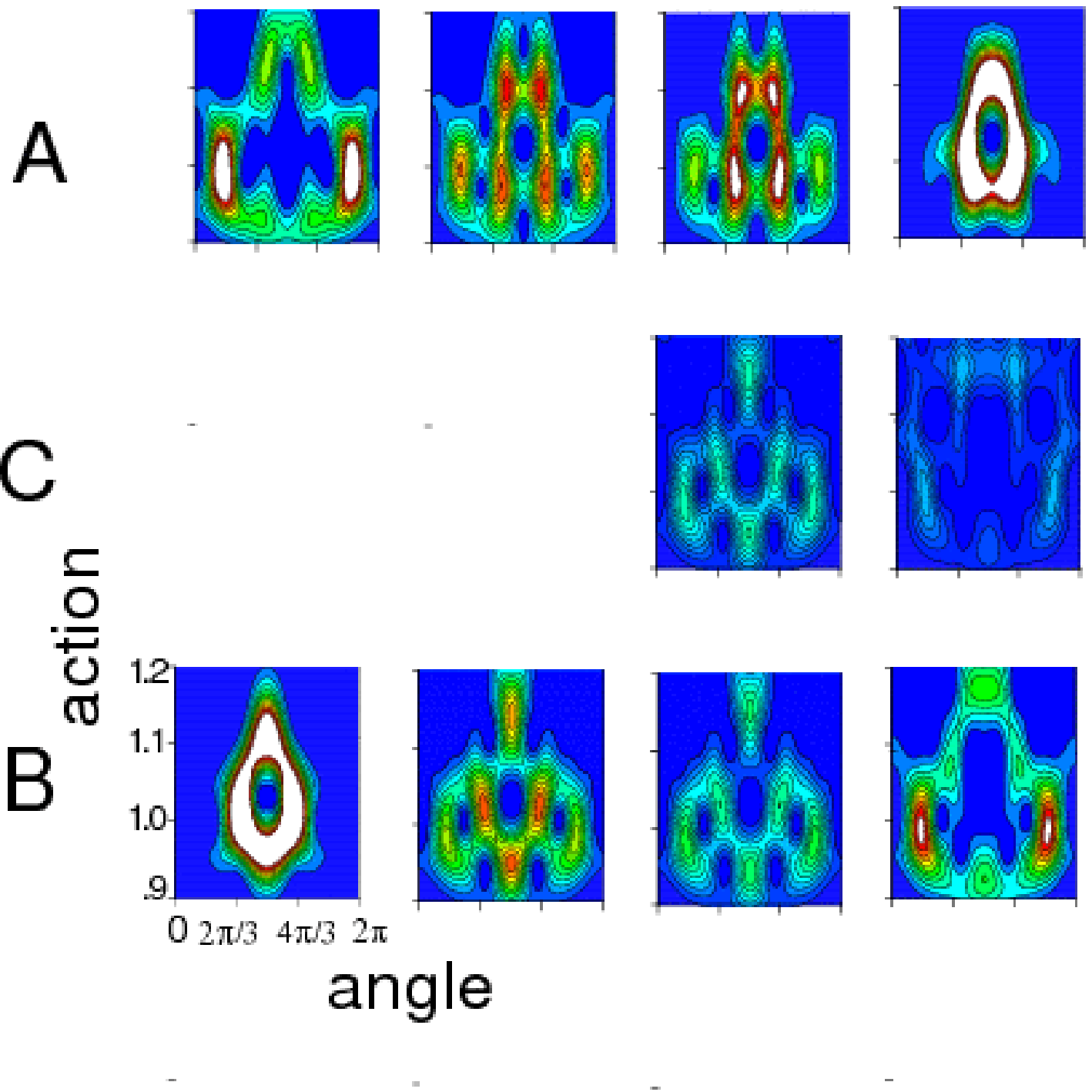,width=0.8\linewidth}
\caption{Detail of Fig. \ref{fig1} showing one of the avoided crossings marked in
Fig. \ref{fig1} and the Husimi functions of the (eigen-)states
undergoing this same avoided crossing: it is a wide crossing where
the lower ``adiabatic" state is made up of two states (B and C):
the third $q=6$ crossing ($k_1=2$, $k_2=8$). The (adiabatic) states
are labeled by letters and numbers indicate the (diabatic)
resonance quantization $k$. Bigger dots mark on the quasienergy
curves the points corresponding to the Husimi functions displayed.
Lighter shades of gray between the level curves mean higher values
of the Husimi functions. The levels of the curves is the same for
all plots; the highest peaks in some of them are, as a result, out
of range and appear as white areas.}
\label{fig9c}
\end{figure}

\begin{figure}[htbp]
\centering\epsfig{file=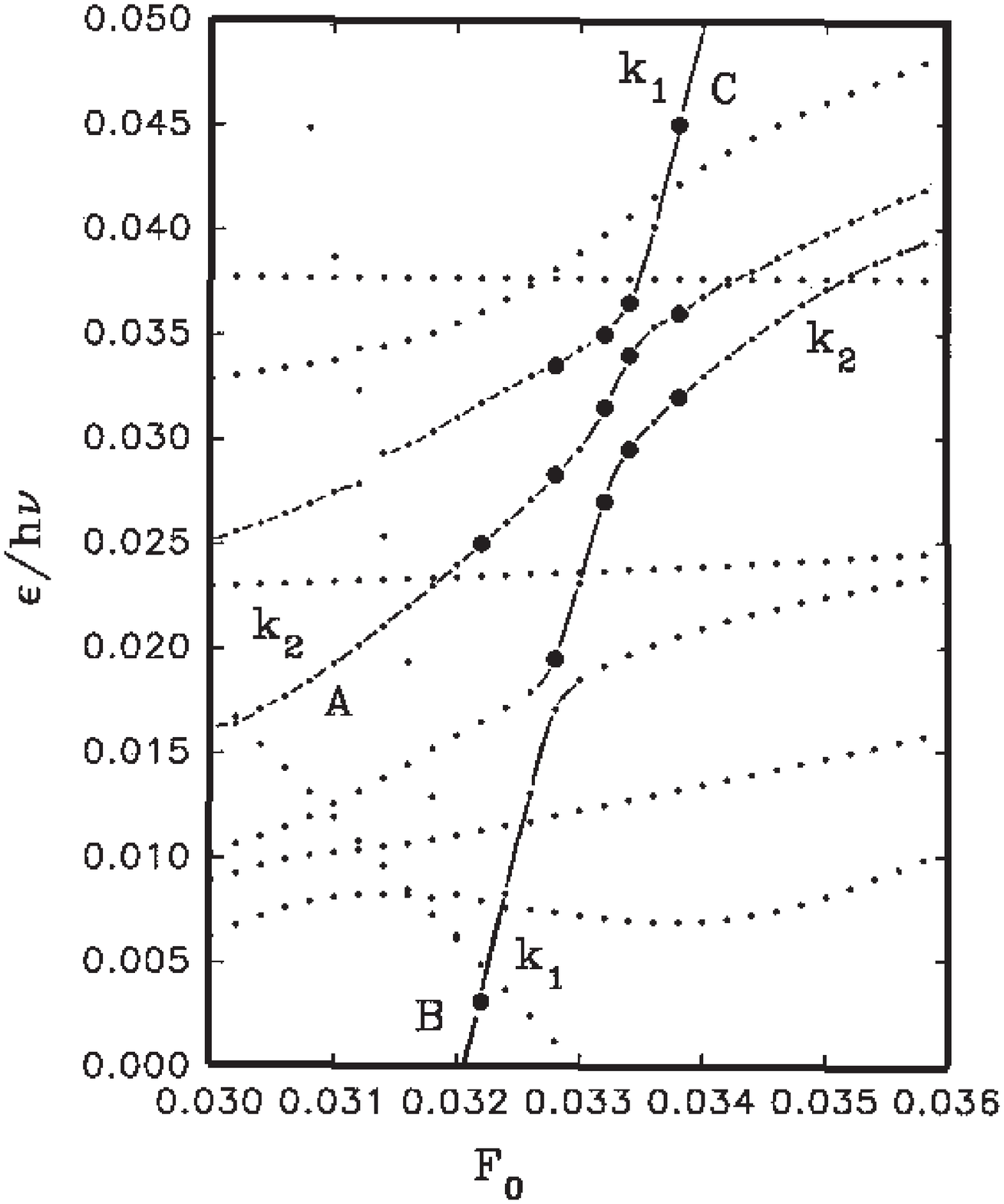,width=0.8\linewidth}
\end{figure}

\begin{figure}[htbp]
\centering\epsfig{file=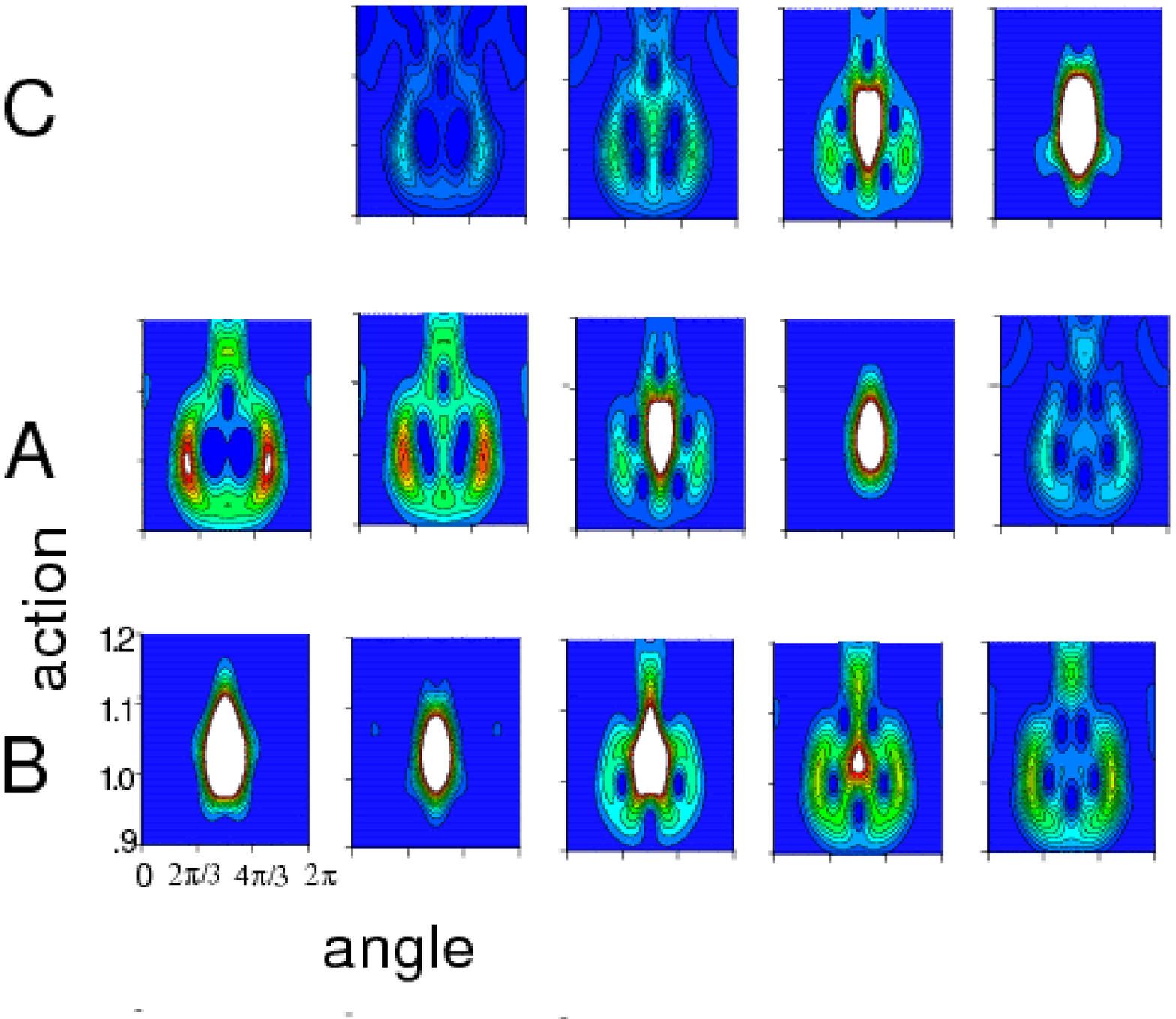,width=0.8\linewidth}
\caption{Detail of Fig. \ref{fig1} showing one of the avoided crossings marked in
Fig. \ref{fig1} and the Husimi functions of the (eigen-)states
undergoing this same avoided crossing: it is a wide crossing where
the upper ``adiabatic" state is made up of two states (A and C):
the first $q=5$ crossing ($k_1=0$, $k_2=5$). The (adiabatic) states
are labeled by letters and numbers indicate the (diabatic)
resonance quantization $k$. Bigger dots mark on the quasienergy
curves the points corresponding to the Husimi functions displayed.
Lighter shades of gray between the level curves mean higher values
of the Husimi functions. The levels of the curves is the same for
all plots; the highest peaks in some of them are, as a result, out
of range and appear as white areas.}
\label{fig9d}
\end{figure}

\begin{figure}[htbp]
\centering\epsfig{file=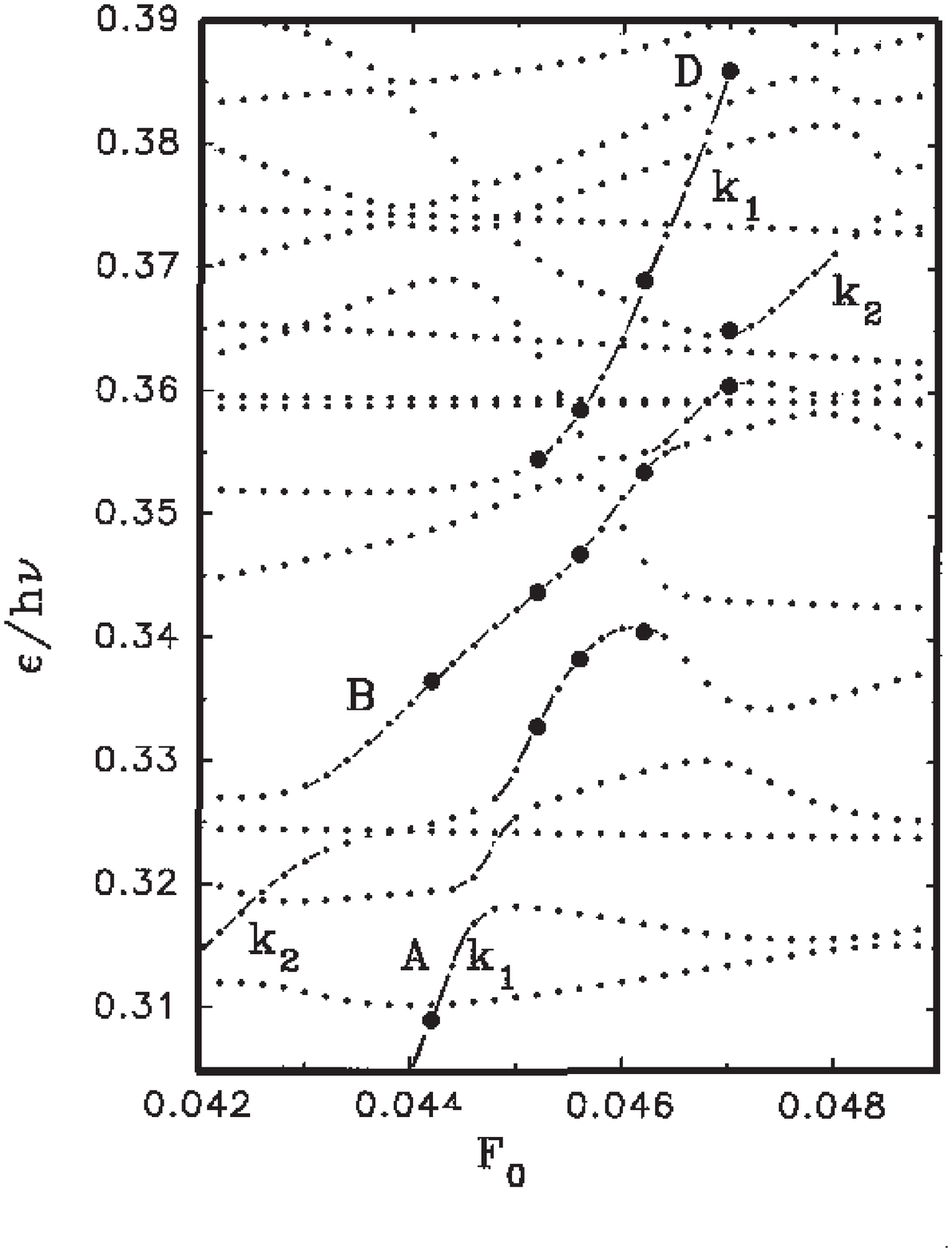,width=0.8\linewidth}
\end{figure}

\begin{figure}[htbp]
\centering\epsfig{file=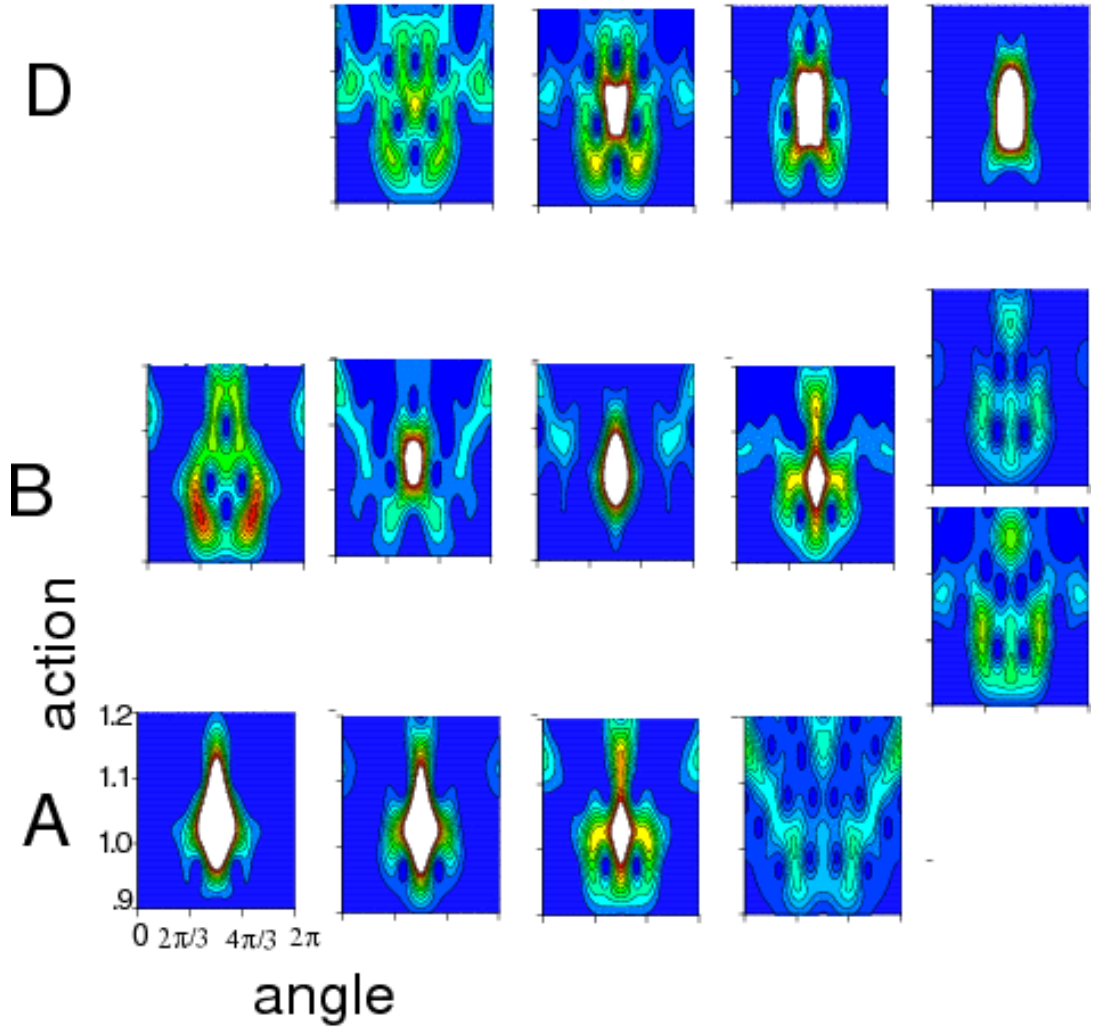,width=0.8\linewidth}
\caption{Detail of Fig. \ref{fig1} showing one of the avoided crossings marked in
Fig. \ref{fig1} and the Husimi functions of the (eigen-)states
undergoing this same avoided crossing: it is a wwide crossing where
a crossing state (B) maintains the same (diabatic) resonance
quantum number $k_2=4$: the first $q=4$ crossing ($k_1=0$,
$k_2=4$). The (adiabatic) states are labeled by letters and numbers
indicate the (diabatic) resonance quantization $k$. Bigger dots
mark on the quasienergy curves the points corresponding to the
Husimi functions displayed. Lighter shades of gray between the
level curves mean higher values of the Husimi functions. The levels
of the curves is the same for all plots; the highest peaks in some
of them are, as a result, out of range and appear as white areas.}
\label{fig9e}
\end{figure}

\begin{figure}[htbp]
\centering\epsfig{file=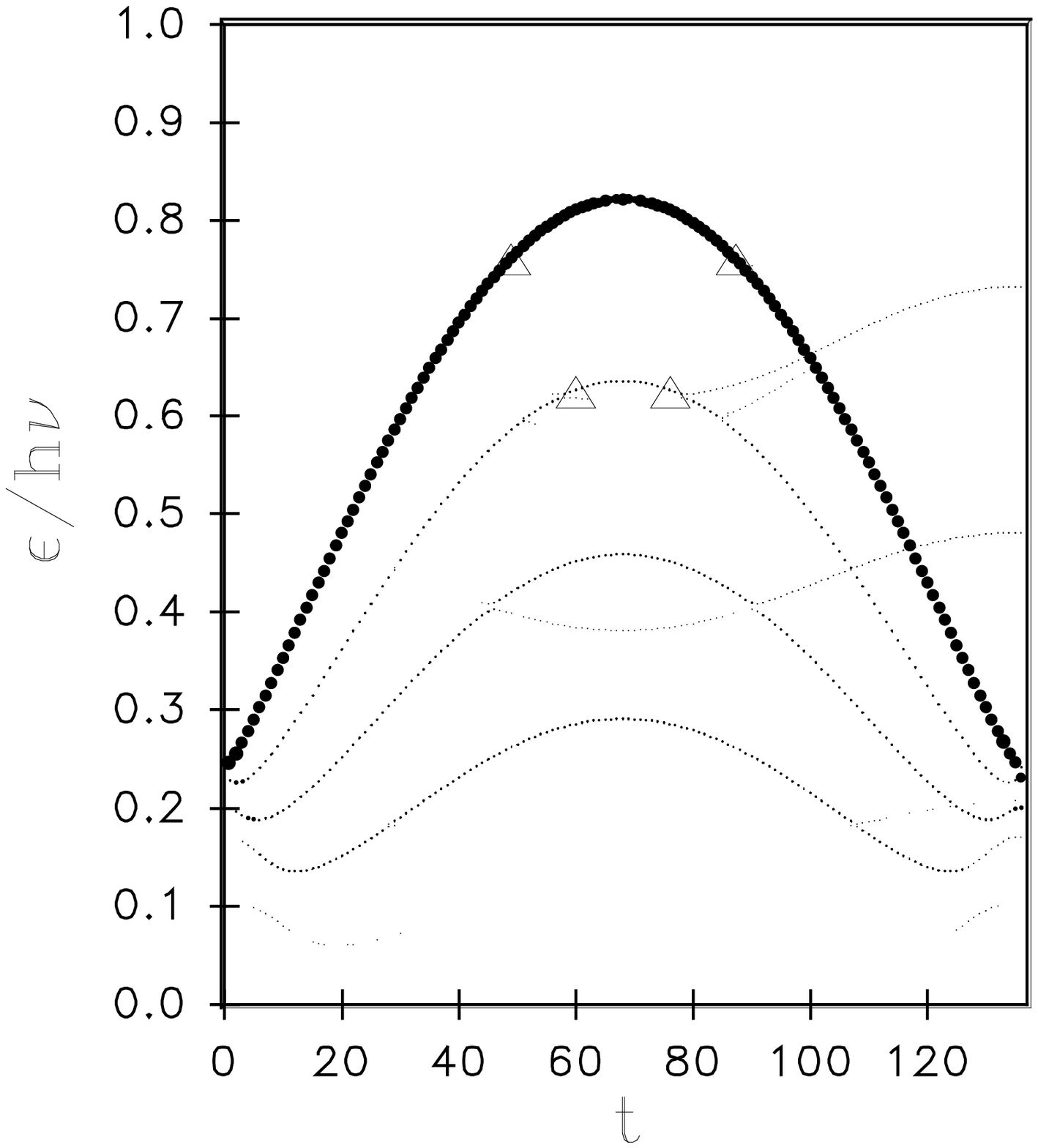,width=0.8\linewidth}
\caption{The quasienergies of the states most significantly populated at
every period of the microwave pulse for the case of Fig.
\ref{fig3a} ($F_0^{max} = 0.025$). The (linear) size of each dot is
proportional to the population on the corresponding level at that
point, the minimum size corresponding to a $0.5\%$ population. The
same symbols as in Fig. \ref{fig1} mark the avoided crossings
related to classical secondary resonance island chains.}
\label{fig10a}
\end{figure}

\begin{figure}[htbp]
\centering\epsfig{file=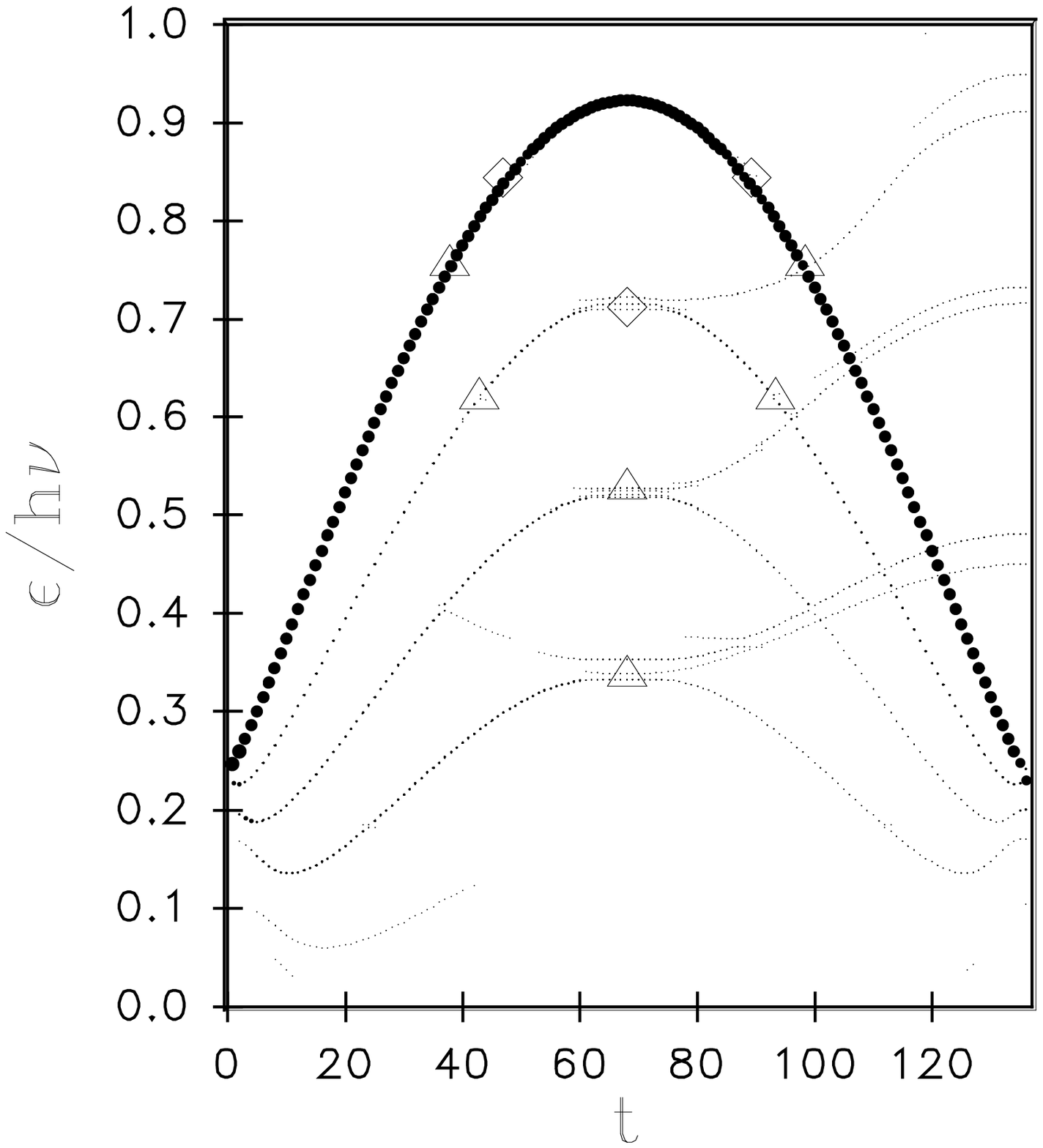,width=0.8\linewidth}
\caption{The quasienergies of the states most significantly populated at
every period of the microwave pulse for the case of Fig.
\ref{fig3b} ($F_0^{max} = 0.029$). The (linear) size of each dot is
proportional to the population on the corresponding level at that
point, the minimum size corresponding to a $0.5\%$ population. The
same symbols as in Fig. \ref{fig1} mark the avoided crossings
related to classical secondary resonance island chains.}
\label{fig10b}
\end{figure}

\begin{figure}[htbp]
\centering\epsfig{file=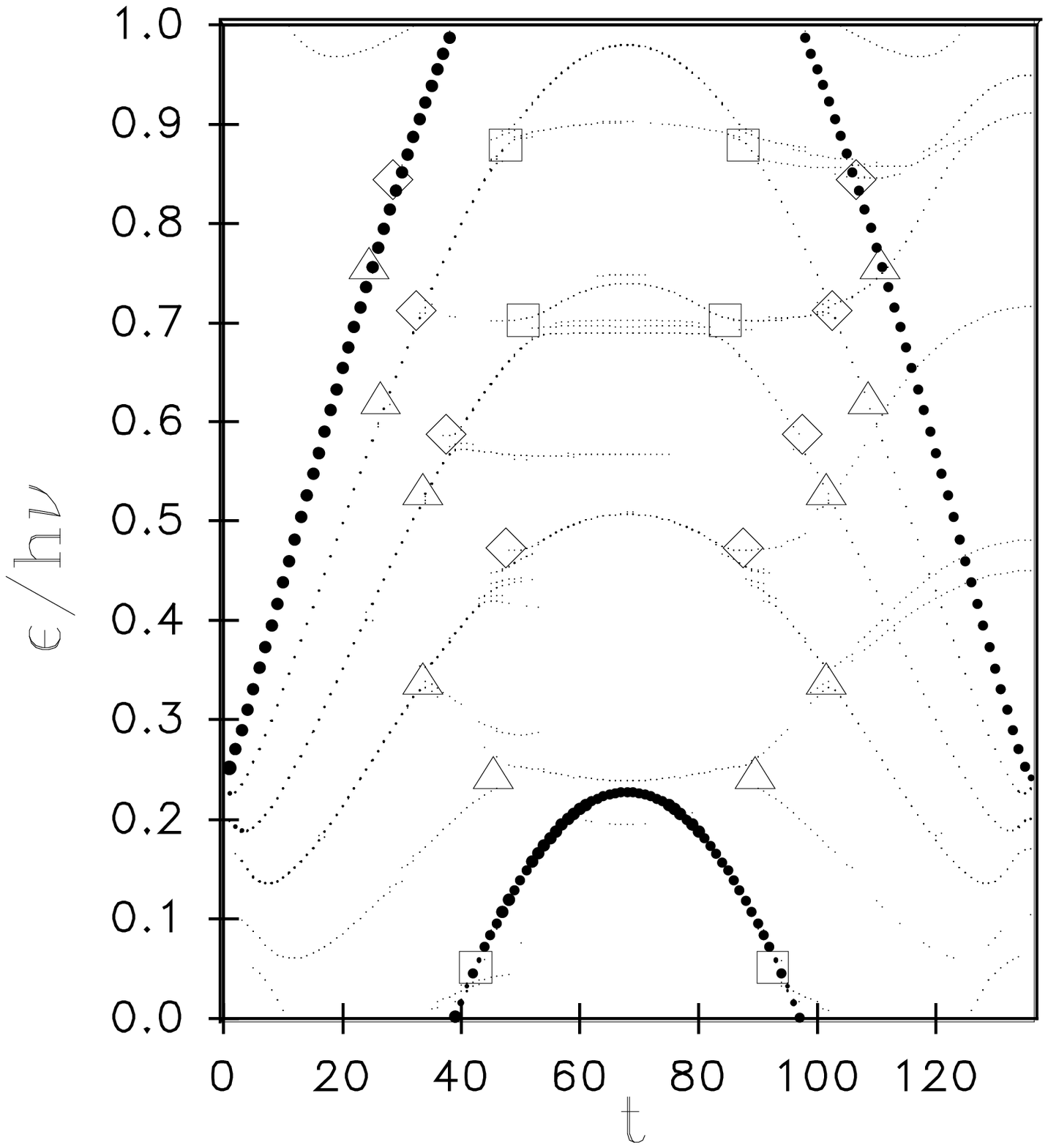,width=0.8\linewidth}
\caption{The quasienergies of the states most significantly populated at
every period of the microwave pulse for the case of Fig.
\ref{fig3c} ($F_0^{max} = 0.041$). The (linear) size of each dot is
proportional to the population on the corresponding level at that
point, the minimum size corresponding to a $0.5\%$ population. The
same symbols as in Fig. \ref{fig1} mark the avoided crossings
related to classical secondary resonance island chains.}
\label{fig10c}
\end{figure}

\begin{figure}[htbp]
\centering\epsfig{file=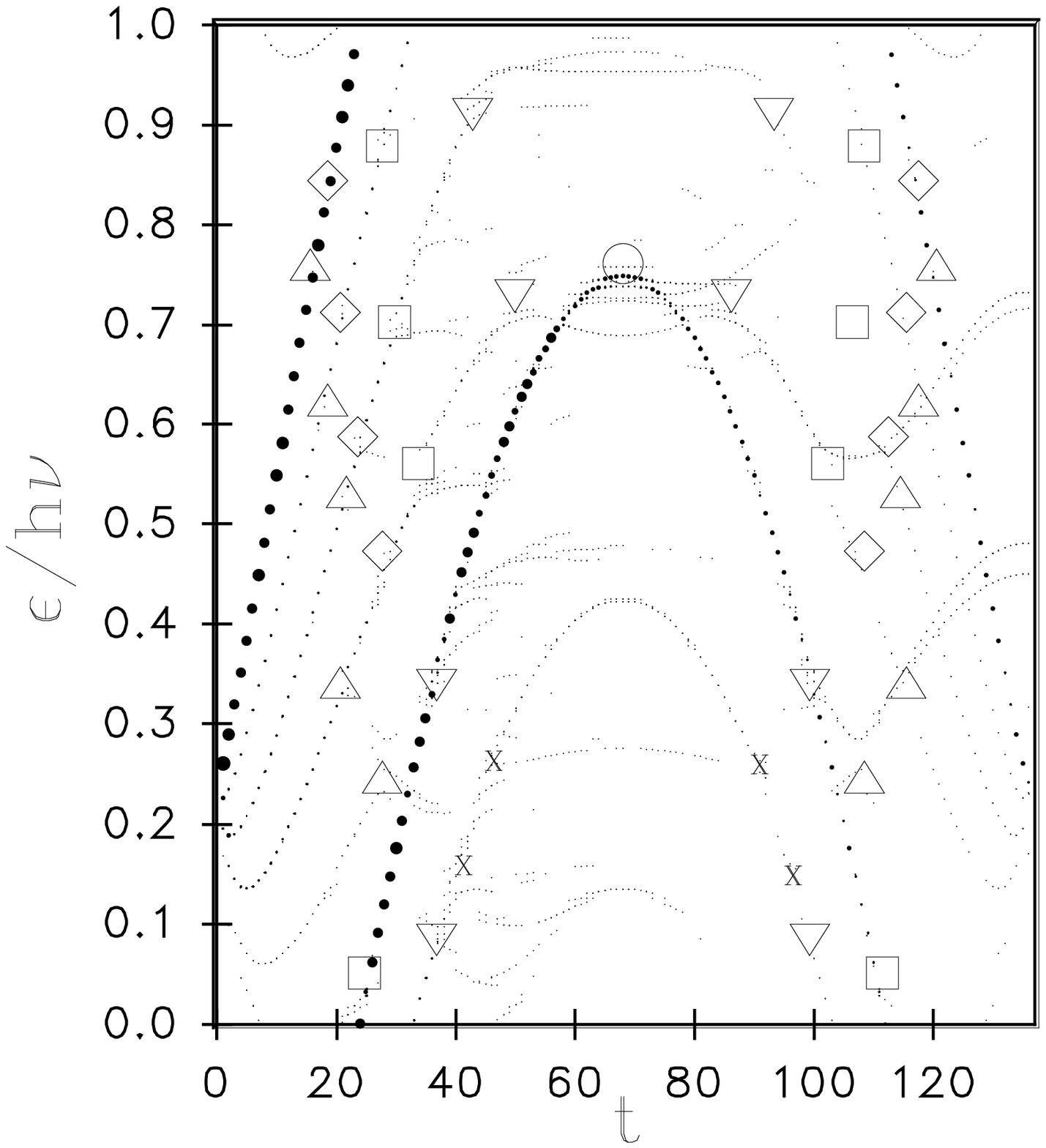,width=0.8\linewidth}
\caption{The quasienergies of the states most significantly populated at
every period of the microwave pulse for the case of Fig.
\ref{fig3d} ($F_0^{max} = 0.061$). The (linear) size of each dot is
proportional to the population on the corresponding level at that
point, the minimum size corresponding to a $0.5\%$ population. The
same symbols as in Fig. \ref{fig1} mark the avoided crossings
related to classical secondary resonance island chains. The crosses
mark crossings with $p=2$ and -from top to bottom- $q=9$, $9$ and
$10$.}
\label{fig10d}
\end{figure}

\begin{figure}[htbp]
\centering\epsfig{file=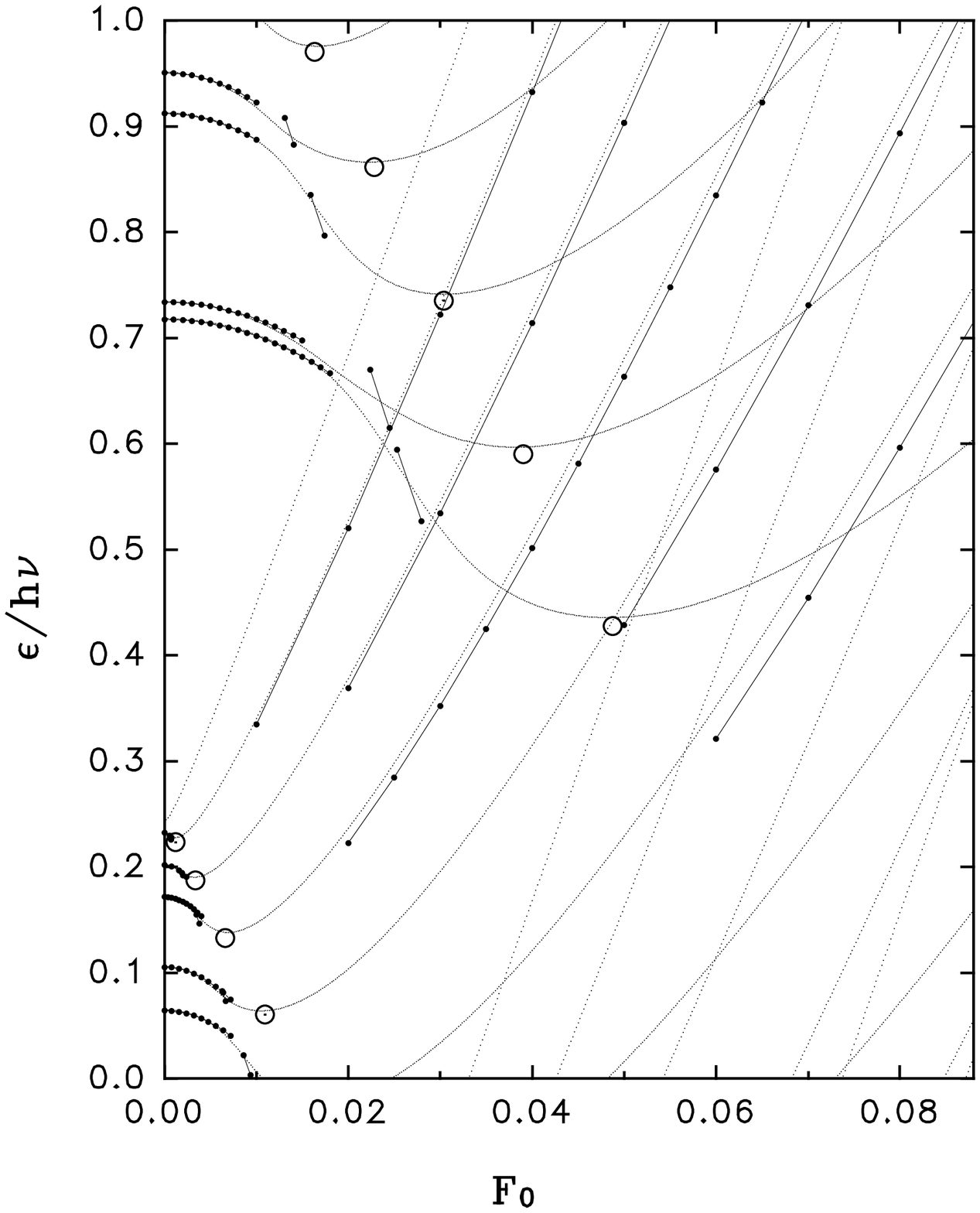,width=0.8\linewidth}
\caption{Comparison of the $k =0-9$ curves obtained from eq. (\ref{eqb14}) using the
off diagonal matrix elements (\ref{eqb13b}) (small dots) with the
classical prediction (big dots) W.K.B. quantized according to the
free atom quantum number $n$ outside the separatrix and according
to the resonance quantum number $k$ inside. Both quantizations are
shown at the separatrix itself, connected by a segment of straight
line. (From Ref. \cite{ref40})}
\label{figb1}
\end{figure}



\begin{thebibliography}{43}
\bibitem{ref24}If the evolution is adiabatic, each point remains at the same resonance action
$J$ and since new actions keep being added by the growth of the
resonance zone, the span in angle of the ensemble decreases. If
instead the evolution is diabatic, the ensemble is not ``squeezed"
but does not expand either. Non-adiabatic evolution will be
intermediate between the two, resulting in partial ``squeezing".
Note that since the resonance action on an invariant (resonance)
curve is the area bounded by the invariant curve itself divided by
$2\pi$, ``squeezing" a resonance torus in the angle variable will
cause it to expand in the action variable. The same conclusion is
reached by applying  Liouville's theorem to the area included
within a resonance invariant curve.
\bibitem{ref26}This also explains a feature of our classical ionization vs. $\omega'_0$ curves
at constant $F_0^{max}$: all curves showed a marked dip at
$\omega'_0\cong1$, but the position in $\omega'_0$ of the minimum
of this dip depended on $F_0^{max}$ in an oscillatory manner. The
point is that the higher is $\omega'_0$, the higher is $F_0^{max}$
corresponding to a given $F_r^{max}$. This means that in a plot at
constant $F_0^{max}$ it is possible for a secondary island chain to
be destroyed at low frequencies but not at high ones, resulting in
lower ionization probabilities for high frequencies and therefore
in the curve having its minimum at $\omega'_0>1$. This will happen
for those values of $F_0^{max}$ close to the onset of a step; for
$F_0^{max}$ close to the center of a plateau we instead expect the
shape of the dip to be determined principally by the primary
resonance. As expected, the minima of the dips are at $\omega'_0=1$
except near the steps where they are at $\omega'_0\gtrsim1$.
\bibitem{ref32}In effect the argument in Ref. \cite{ref25} is that no single island in the
secondary chains is big enough to significantly support a quantum
state. It is equivalent to requiring quantum multilevel (involving
more than two levels at a time) interactions relating to a $p/q$
classical resonance zone \cite{ref40} and it is a more stringent
condition than the one we require, since even in the case of
overlapping crossing regions the quantum dynamics may still be a
two level one: if $q\neq1$, the pairs of states coupled by the
overlapping crossing zones may have no state in common, even if the
crossing regions are related to the same classical resonance zone.
As a simple example let us consider the $p/q=1/3$ crossing in Fig.
\ref{fig8}: with increasing $F$, the first crossing we encounter is
the $(k_1,k_2)=(0,3)$ one (just to the right the classical
$p/q=1/3$ bifurcation field); the following one is the $(1,4)$ one,
followed by (out of the picture) the $(2,5)$ one and only then by
the $(3,6)$ one which again has $k=3$ as a level taking part in the
crossing. To be able to have one quantum multilevel (involving more
than two levels at a time) interaction relating to a p/q classical
resonance zone we therefore need at least $q+1$ crossing regions to
overlap at the same $F$. Based on the $q$-fold structure our Husimi
functions at a crossing in Figs. \ref{fig9a}-\ref{fig9e}, we prefer
to use the size of the whole chain, instead of that of a single
island. Moreover, from Figs. \ref{fig9a}-\ref{fig9e} it is clear
that the structure responsible for the steps is not a single
secondary island, but the whole chain of  them. Finally, we find
fault with the argument (often adduced in support of considering
the area of a single island of the chain) that, since the system
(at constant field strength) is time periodic, all the islands of a
chain belong to the same vortex tube in the extended phase space
${\mathcal P}
=\{I, \theta, H, t\in(0,T)\}$ and therefore only one section (a
single island) of the tube should be considered when evaluating how
many quantum states (of area $h$) the classical chain can support.
Our objection is twofold: first, the quantities to compare would in
this case be the volume of the vortex tube in $\mathcal{P}$ and
$h^2$. Second, as pointed out in Ref. \cite{ref45}, quantization of
a single vortex tube would result in a $qT$ periodic wavefunction,
which is not a Floquet eigenstate.
\bibitem{ref47}A comparable classical spread can be inferred for example from the first snapshot
of Figure \ref{fig4b} where the ensemble is spread over an
approximately triangular region of (rescaled) area $A\cong
0.14(2\pi0.83)/2$ corresponding to a spread $\Delta k =
n_0A/2\pi\cong 4$.
\end{thebibliography}
\end{document}